\documentclass[prb,preprint,tightenlines,
  groupedaddress,amsmath,amssymb,amsfonts]{revtex4}

\usepackage{bm}
\usepackage{graphicx}

\newcommand{\inside}{<}
\newcommand{\outside}{>}
\newcommand{\defeq}{=}
\newcommand{\optimal}{\star}

\begin{document}

  \preprint{EKM-TP3/22-01}
  
  \title{Exact~analytic~results for the Gutzwiller~wave~function
    with~finite~magnetization}

  \author{Marcus Kollar}
  \email[E-Mail: ]{Marcus.Kollar@physik.uni-augsburg.de}
  \altaffiliation[Address after December~1, 2001: ]{
    In\-sti\-tut f\"{u}r  Theore\-ti\-sche Phy\-sik,
    Jo\-hann-Wolf\-gang-Goethe-Uni\-ver\-si\-t\"{a}t Frank\-furt,
    Ro\-bert-Mayer-Stra\ss{}e~8-10,
    D-60054~Frankfurt am~Main,
    Germany.}

  \author{Dieter Vollhardt}
  \email[E-Mail: ]{Dieter.Vollhardt@physik.uni-augsburg.de}
  \affiliation{Theoretische Physik III,
    Elektronische Korrelationen und Magnetismus,
    Institut f\"{u}r Physik, Universit\"{a}t Augsburg,
    D-86135 Augsburg, Germany}

  \date{November 2, 2001}

  \begin{abstract}
    We present analytic results for ground-state properties of
    Hubbard-type models in terms of the Gutzwiller variational wave
    function with non-zero values of the magnetization $m$.  In
    dimension $D$ $=$ $1$ approximation-free evaluations are made
    possible by appropriate canonical transformations and an analysis
    of Umklapp processes. We calculate the double occupation and the
    momentum distribution, as well as its discontinuity at the Fermi
    surface, for arbitrary values of the interaction parameter $g$,
    density $n$, and magnetization $m$.  These quantities determine
    the expectation value of the one-dimensional Hubbard Hamiltonian
    for any symmetric, monotonically increasing dispersion
    $\epsilon_k$.  In particular for nearest-neighbor hopping and
    densities away {}from half filling the Gutzwiller wave function is
    found to predict ferromagnetic behavior for sufficiently large
    interaction $U$.
  \end{abstract}


  \maketitle

  \section{Introduction}
  \label{sec:introduction}
  
  Quantum-mechanical many-body problems can almost never be solved
  exactly. In this situation variational wave functions have proved to
  be particularly useful. Although they describe correlations among
  the particles only in an approximate way, they have the advantage of
  being explicit and physically intuitive. In particular, they allow
  for investigations even when standard perturbation theory is not
  applicable, or is untractable.
  
  Variational wave functions can, for example, be obtained by applying
  a suitably chosen correlation operator (e.~g., the interaction part
  of the Hamiltonian under investigation) to a simple one-particle
  wave function.  For the one-band Hubbard
  model\cite{Hubbard63,Kanamori63,Gutzwiller63a+64a+65a}
  \begin{align}
    \hat{H}
    =
    \sum_{\bm{k}\sigma}
    \epsilon_{\bm{k}}
    \hat{a}_{\bm{k}\sigma}^{+}\hat{a}_{\bm{k}\sigma}^{\phantom{+}}
    +
    U
    \sum_{i}
    \hat{n}_{i\uparrow}\hat{n}_{i\downarrow}
    ,\,\label{hamiltonian}
  \end{align}
  which is often used as an effective model to understand electronic
  correlation phenomena like itinerant ferromagnetism in transition
  metals, high-temperature superconductivity and the Mott-Hubbard
  metal-insulator transition, the simplest projected wave function is
  the Gutzwiller wave function (GWF)\cite{Gutzwiller63a+64a+65a}
  \begin{align}
    |\Psi_{\text{G}}\rangle
    =
    g^{\sum_{i}\hat{D}_{i}}|\Phi_{0}\rangle
    =
    \prod_{i}\left[ 1-(1-g)\hat{D}_{i}\right]|\Phi_{0}\rangle
    ,\,\label{ansatz}
  \end{align}
  where $g$ is a variational parameter (usually $0\leq g\leq1)$,
  $\hat{D}_{i}=\hat{n}_{i\uparrow}\hat{n}_{i\downarrow}$ is the
  operator for double occupation at lattice site $i$, and the starting
  wave function $|\Phi_{0}\rangle$ is a product state of spin-up and
  spin-down Fermi seas
  \begin{align}
    |\Phi_{0}\rangle
    =
    \prod_{\stackrel{\scriptstyle
        \bm{k}\sigma}{\epsilon_{\bm{k}}\leq\epsilon_{\text{F}\sigma}}}
    \hat{a}_{\bm{k}\sigma}^{+}
    \;|\text{vac}\rangle
    .\,\label{startwf}
  \end{align}
  Using the GWF one may, in principle, calculate expectation values of
  any operator $\hat{A}$ as $\langle\hat{A}\rangle_{\text{G}}$ $=$
  $\langle\Psi_{\text{G}}|\hat{A}|\Psi_{\text{G}}\rangle /
  \langle\Psi_{\text{G}}|\Psi_{\text{G}}\rangle$. By the variational
  principle the energy expectation value
  $\langle{\hat{H}}\rangle_{\text{G}}$ is an upper bound for the true
  ground-state energy of $\hat{H}$.
  
  The properties and quality of the GWF have been subject of detailed
  investigations (for an early review see
  Ref.~\onlinecite{Vollhardt90a}). A diagrammatic theory for the
  calculation of expectation values in terms of the GWF, valid in
  arbitrary dimensions $D$, was formulated by Metzner and
  Vollhardt.\cite{Metzner87a+88a} (Ref.~\onlinecite{Metzner87a+88a} is
  hereafter referred to as MV.)  In particular, for systems without
  net magnetic polarization $m$ $=$ $n_{\uparrow}-n_{\downarrow}=0$
  (i.~e., particle densities $n_{\sigma}=n/2$, with
  $k_{\text{F}\uparrow}=k_{\text{F}\downarrow}$ in
  Eq.~(\ref{startwf})) they calculated the momentum distribution
  $n_{\bm{k}\sigma}$ $=$ $\langle\hat{a}_{\bm{k}\sigma}^{+}
  \hat{a}_{\bm{k}\sigma}^{\phantom{+}}\rangle_{\text{G}}$ and the
  double occupation $d$ $=$
  $\langle{\sum_{i}\hat{D}{_{i}}}\rangle_{\text{G}}/L$ analytically in
  $D=1$ for all values of $g$ and $n$, where $L$ is the number of
  lattice sites. The analytic calculation of correlation functions, in
  particular of the spin-spin correlation function, by Gebhard and
  Vollhardt\cite{Gebhard87a+88a} showed that in the non-magnetic case,
  for $U\rightarrow \infty$, the results obtained with the GWF are in
  very good agreement with exact analytic and numerical results for
  the antiferromagnetic Heisenberg chain. In fact,
  Haldane\cite{Haldane88a} and Shastry\cite{Shastry88a} discovered
  that the GWF at $g$ $=$ $0$ is the \emph{exact} ground state wave
  function of the antiferromagnetic Heisenberg chain with $1/r^2$
  exchange.  Results were also obtained in dimensions $D=1,2,3$ using
  numerical techniques\cite{Yokoyama87a,Yokoyama97a} and finite orders
  of perturbation theory.\cite{Gulacsi91a,Gulacsi93a} Within the
  diagrammatic approach of MV it also became possible to derive the
  well-known Gutzwiller approximation in the limit of infinite spatial
  dimensions ($D=\infty$).\cite{Metzner89a} Comprehensive
  investigations in this limit were made possible by the approach of
  Gebhard,\cite{Gebhard90a} which allows for explicit evaluations of
  expectation values for arbitrary starting wave functions
  $|\Phi_{0}\rangle$ (including ones with broken symmetry) and
  facilitates the expansion in $1/D$ around $D=\infty$. This approach
  was also extended to multi-band Hubbard models;\cite{Buenemann98a}
  recently that method was combined with density functional theory,
  and applied to ferromagnetic transition metals.\cite{Weber01a} The
  Gutzwiller approximation also describes a correlation-induced
  transition {}from metal to insulator, the Brinkman-Rice
  transition.\cite{Brinkman70a} We recently investigated the effect of
  correlated hopping, which for the GWF can be calculated in terms of
  $n_{\bm{k}\sigma}$ and $d$ in any dimension, on this
  transition.\cite{Kollar00a}
  
  Up to now the analytic calculation of expectation values in $D=1$
  was limited to the unmagnetized paramagnetic phase. In view of the
  renewed interest in the microscopic foundations of metallic
  ferromagnetism (see Ref.~\onlinecite{Vollhardt99a} for a review), it
  is desirable to perform such evaluations also for the GWF with {\em
    non-zero} magnetization ($m\neq0$).  In this paper we show that,
  in spite of formidable technical complications, it is indeed
  possible to evaluate such expectation values even for finite
  magnetization.  {}From suitable canonical transformations we obtain
  diagrammatic relations and reduce the expectation values for
  $m\neq0$ to those for $m$ $=$ $0$. Thereby it becomes possible to
  calculate the double occupation $d$ and the momentum distribution
  $n_{k\sigma}$ for arbitrary values of the correlation parameter $g$,
  density $n$, and magnetization $m$.  Furthermore we derive an
  expression for $n_{k\sigma}$ in closed form, which was not available
  up to now even for zero magnetization.  These quantities determine
  the energy expectation value and thus the optimal variational
  parameter and spontaneous magnetization.
  
  The paper is structured as follows: In Sec.~\ref{sec:diagrammatics}
  the diagrammatic formulation is used to derive diagrammatic
  relations {}from canonical transformations, valid in arbitrary
  dimensions. The evaluation of expectation values is derived in
  Sec.~\ref{sec:evaluation}.  The resulting magnetic phase diagram for
  the Hubbard model in $D=1$ is presented in Sec.~\ref{sec:results}.
  The conclusion in Sec.~\ref{sec:conclusion} closes the presentation.

  \section{Diagrammatic formulation in arbitrary dimension $D$}
  \label{sec:diagrammatics}

  \subsection{General formalism}
  
  The double occupation $d(g,n,m)$ and the momentum distribution
  $n_{\bm{k}\sigma}(g,n,m)$ of the GWF are required for the
  calculation of the variational energy, $E_{\text{G}}$ $=$
  $\langle{\hat{H}}\rangle_{\text{G}}/L$.  Another quantity of
  interest is the discontinuity $q_\sigma$ of $n_{\bm{k}\sigma}$ at
  the Fermi surface, $q_{\sigma}(g,n,m)$ $=$
  $n_{\bm{k}_{\text{F}\sigma}^{-}\sigma}(g,n,m)$ $-$
  $n_{\bm{k}_{\text{F}\sigma}^{+}\sigma}(g,n,m)$.\cite{note-on-q} The
  rules for the diagrammatic expansion of these expectation values in
  powers of $(g^2-1)$ were developed by MV, with the result
  \begin{subequations}\label{entwicklungen}
    \begin{align}
      d(g,n,m)
      &=
      g^2\sum_{p=1}^{\infty}(g^2-1)^{p-1}c_{p}(n,m)
      ,\,\label{doppelbes}
      \\
      n_{\bm{k}\sigma}(g,n,m)
      &=
      (1-(1-g)^2{n}_{\bar{\sigma}})
      n_{\bm{k}\sigma}^{0}
      +
      \frac{1-(1-g^2) n_{\bm{k}\sigma}^{0}}{(1+g)^2}
      \sum_{p=2}^{\infty}(g^2-1)^{p} f_{p\sigma}(\bm{k},n,m)
      \label{kraumbes}
    \end{align}
  \end{subequations}
  where $n_{\bm{k}\sigma}^{0}$ $=$ $n_{\bm{k}\sigma}(g=1,n,m)$.  The
  functions $c_{p}(n,m)$ and $f_{p\sigma}(\bm{k},n,m)$ can be
  represented by Feynman diagrams corresponding to those of the energy
  and the Greens function, respectively, of a $\phi^{4}$ theory. For
  later convenience we define
  \begin{align}
    f_{1\sigma}
    \defeq
    -n_{\bar{\sigma}}n_{\bm{k}\sigma}^{0}
    ,\;\;
    f_{0\sigma}
    \defeq
    n_{\bm{k}\sigma}^{0}
    ,\;\;
    c_{0}
    \defeq
    -\frac{n-|m|}{2}
    .\,\label{f0h0c0def}
  \end{align}
  The diagrams for $c_{p}(n,m)$ can be obtained {}from those for
  $f_{p\sigma}(\bm{k},n,m)$ by connecting the two external vertices
  (see MV):
  \begin{align}
    c_{p}(n,m)
    =
    -
    \frac{1}{L}\sum_{\bm{k}}
    n_{\bm{k}\sigma}^{0}
    f_{p\sigma}(\bm{k},n,m)
    ,\;\;p\geq1
    .\,\label{cfrelationin}
  \end{align}
  This equation yields sum rules\cite{Hashimoto85a} for the density of
  particles inside, $n_{\sigma}^{\inside}$, or outside of the Fermi
  surface, $n_{\sigma}^{\outside}$, namely
  \begin{align}
    n_{\sigma}^{\inside}
    &\defeq
    n_{\sigma}-n_{\sigma}^{\outside}
    \defeq
    \frac{1}{L}\sum_{\bm{k}}
    n_{\bm{k}\sigma}^{0} n_{\bm{k}\sigma}
    ,\,\label{n_in-def}
    \\
    n_{\uparrow}^{\outside}
    &=
    n_{\downarrow}^{\outside}
    =
    \frac{1-g}{1+g}\left(\frac{n^2-m^2}{4}-d(g,n,m)\right)
    \!.\,\label{sumrule}
  \end{align}
  It is sometimes useful to remove the diagrams $c_p$ {}from
  $f_{p\sigma}$ and thus define
  \begin{align}
    h_{p\sigma}(\bm{k},n,m)
    =
    f_{p\sigma}(\bm{k},n,m)-c_{p-1}(n,m),
    \;\;p\geq1
    .\,\label{fhcdef}
  \end{align}
  Note that our definitions in Eqs.~(\ref{f0h0c0def}) and
  (\ref{fhcdef}) differ slightly {}from MV.

  \subsection{Canonical transformations}\label{subsec:trafos}
  
  For the relations to be discussed next the hopping amplitude
  $t_{ij}$ is assumed to be non-zero only for hopping between sites
  $i$ and $j$ on different sublattices $A$ and $B$. In the next
  chapter we will see, however, that in dimension $D=1$ this
  requirement can be dropped. The simplest canonical transformation is
  the interchange of spin indices $\uparrow$ and $\downarrow$ which
  implies
  \begin{subequations}
    \label{negmagn}
    \begin{align}
      \label{negmagndocc}
      d(g,n,m)
      &=
      d(g,n,-m)
      ,\\
      n_{\bm{k}\sigma}(g,n,m)
      &=
      n_{\bm{k}\bar{\sigma}}(g,n,-m)
      .\,\label{negmagnkocc}
    \end{align}
  \end{subequations}
  Furthermore, a particle-hole transformation for both spins yields
  (MV)
  \begin{subequations}
    \begin{align}
      d(g,n,m)
      &=
      d(g,2-n,-m)+n-1
      ,\,\label{bigdensdocc}\\
      n_{\bm{k}\sigma}(g,n,m)
      &=
      1-n_{\bm{Q}-\bm{k}\sigma}(g,2-n,-m)
      .\,\label{bigdenskocc}
    \end{align}
    \label{bigdens}
  \end{subequations}
  Here $\bm{Q}$ is a vector in the first Brillouin zone with
  $e^{\text{i}\bm{Q}\cdot\bm{R}}=\pm1$ for a lattice vector $\bm{R}\in
  A$, $B$, respectively; $\bm{Q}=(\pi/a,\pi/a,\ldots\pi/a)$ for a
  hypercubic lattice with spacing $a$. {}From Eqs.~(\ref{negmagnkocc})
  and (\ref{bigdenskocc}) it follows that the discontinuity at the
  Fermi surface obeys\cite{note-on-q}
  \begin{align}
    q_{\sigma}(g,n,m)
    =
    q_{\bar{\sigma}}(g,n,-m)
    =
    q_{\bar{\sigma}}(g,2-n,m)
    .\,\label{qdisctrafo}
  \end{align}
  Note that in particular $q_{\uparrow}=q_{\downarrow}$ for $n$ $=$
  $1$.
  
  For densities $0$ $\leq$ $n$ $\leq$ $2$ the magnetization is in the
  range $|m|$ $\leq$ $\min(n,2-n)$.  In view of Eqs.~(\ref{negmagn}),
  (\ref{bigdens}), and (\ref{qdisctrafo}) we limit ourselves {}from
  now on to $0\leq m< n\leq1$. Therefore $\sigma$ $=$ $\uparrow$ will
  be referred to as ``majority spin'' and $\sigma$ $=$ $\downarrow$ as
  ``minority spin.''  Note that the case $n$ $=$ $m$ (the fully
  polarized state without doubly occupied sites) can be obtained
  {}from the uncorrelated case $g$ $=$ $1$.
  
  Performing a particle-hole transformation for $\uparrow$-operators
  only,\cite{Gebhard87a+88a}
  \begin{align}
    \hat{c}_{i\uparrow}^{\prime}
    =
    (-1)^{i}\hat{c}_{i\uparrow}^{+}
    ,\;\;
    \hat{c}_{i\downarrow}^{\prime}
    =
    \hat{c}_{i\downarrow}^{\phantom{+}}
    ,\,\label{trafo}
  \end{align}
  one may derive the following identities for $d(g,n,m)$ and
  $n_{\bm{k}\sigma}(g,n,m)$:
  \begin{subequations}
    \begin{align}
      d(g,n,m)
      &=
      \frac{n-m}{2}-d(g^{-1},1-m,1-n)
      .\,\label{drelation}
      \\
      n_{\bm{k}\uparrow}(g,n,m)
      &=
      1-n_{\bm{Q}-\bm{k}\uparrow}(g^{-1},1-m,1-n)
      ,\,\label{kuprelation}
      \\
      n_{\bm{k}\downarrow}(g,n,m)
      &=
      n_{\bm{k}\downarrow}(g^{-1},1-m,1-n)
      .\,\label{kdownrelation}
    \end{align}
    \label{ginvrelations}
  \end{subequations}
  For the uncorrelated case ($g$ $=$ $1$) we have in particular
  \begin{subequations}\label{nonintrelations}
    \begin{align}
      n_{\bm{k}\uparrow}^0(n,m)
      &=
      1-n_{\bm{Q}-\bm{k}\uparrow}^0(1-m,1-n)
      ,\,\label{nonintkuprelation}
      \\
      n_{\bm{k}\downarrow}^0(n,m)
      &=
      n_{\bm{k}\downarrow}^0(1-m,1-n)
      .\,\label{nointkdownrelation}
    \end{align}
  \end{subequations}
  The relations in Eq.~(\ref{nonintrelations}) express a property of
  the starting wave function [Eq.~(\ref{startwf})] and can also be
  derived directly {}from the fact that hopping occurs only between
  $A$ and $B$ sublattices.

  \subsection{Diagrammatic relations}\label{subsec:diagrels}
  
  We now derive diagrammatic relations for $c_{p}$ and $f_{p\sigma}$
  {}from the identities in Eqs.~(\ref{ginvrelations}) and
  (\ref{nonintrelations}). The following equations are valid for all
  $p\geq0$ and $0\leq m\leq n\leq1$ unless noted otherwise.

  \subsubsection{Double occupation}\label{subsubsec:diagdocc}
  
  {}From Eq.~(\ref{doppelbes}) we obtain
  \begin{align}
    d(g^{-1},1-m,1-n)
    =
    \sum_{p=0}^{\infty}
    (g^2-1)^{p}(-1)^{p}
    \left(
      \frac{1}{g^2}
    \right)^{p+1}
    c_{p+1}(1-m,1-n)
    .\,\label{dprime}
  \end{align}
  Our goal is to equate coefficients of powers of $(g^2-1)$. To this
  end we make use of the expansion
  \begin{align}
    \left(\frac{1}{g^2}\right)^{p+1}=\sum_{r=0}^{\infty}
    \binom{-p-1}{r}
    (g^2-1)^{r}
    .\,\label{ginvexpansion}
  \end{align}
  We then obtain {}from Eq.~(\ref{drelation})
  \begin{align}
    d(g,n,m)
    =
    \frac{n-m}{2}
    -
    \sum_{p=0}^{\infty}(g^2-1)^{p}\,(-1)^{p}
    \sum_{r=0}^{p}
    \binom{p}{r}
    c_{r+1}(1-m,1-n)
    ,\,\label{dtransformiert}
  \end{align}
  while {}from Eq.~(\ref{doppelbes}) we have
  \begin{align}
    d(g,n,m)
    =
    c_{1}(n,m)
    +
    \sum_{p=1}^{\infty}(g^2-1)^{p}\,
    \left(
      c_{p}(n,m)+c_{p+1}(n,m)
    \right)
    \!.\,\label{dumsummiert}
  \end{align}
  We are thus led to the relation
  \begin{align}
    c_{p}(n,m)+c_{p+1}(n,m)
    =
    (-1)^{p+1}
    \sum_{r=0}^{p}
    \binom{p}{r}
    c_{r+1}(1-m,1-n)
    .\,\label{erstecrelation}
  \end{align}
  Now we employ the \emph{binomial inversion formula}\cite{Wilf94a}
  \begin{alignat}{3}
    &&
    a_{p}&=
    (-1)^{p}\sum_{q=0}^{p}
    \binom{p}{q}
    b_{q}
    &&
    \text{for all }p\geq0
    \nonumber\\
    &\Leftrightarrow\;\;\;&
    b_{p}&=
    (-1)^{p}\sum_{q=0}^{p}
    \binom{p}{q}
    a_{q}
    &\;\;&
    \text{for all }p\geq0
    ,\,\label{binominv}
  \end{alignat}
  which is valid for arbitrary $a_{p}$ and $b_{p}$. When applied to
  Eq.~(\ref{erstecrelation}) it yields
  \begin{align}
    c_{p}(n,m)
    &=
    (-1)^{p}
    \sum_{r=0}^{p}
    \binom{p}{r}
    c_{r}(1-m,1-n)
    .\,\label{crelation}
  \end{align}
  We stress that the relations in Eqs.~(\ref{erstecrelation}) and
  (\ref{crelation}) are valid in arbitrary dimensions for lattices
  with hopping between $A$ and $B$ sublattices only.  In the next
  section this equation will be used to calculate $c_{p}$ in $D=1$.

  \subsubsection{Momentum distribution}\label{subsubsec:diagkocc}
  
  An analogous procedure is used to derive relations for
  $f_{p\sigma}$.  We define the abbreviations
  \begin{subequations}\label{Fdefs}
    \begin{align}
      F_{p\sigma}(\bm{k},n,m)
      &\defeq
      f_{p+2\sigma}(\bm{k},n,m)
      +n_{\bm{k}\sigma}^{0}f_{p+1\sigma}(\bm{k},n,m)
      ,\,\label{Fdefsigma}
      \\
      \bar{F}_{p\uparrow}(\bm{k},n,m)
      &\defeq
      F_{p\uparrow}(\bm{Q}-\bm{k},1-m,1-n)
      ,\,\label{Fdefup}
      \\
      \bar{F}_{p\downarrow}(\bm{k},n,m)
      &\defeq
      F_{p\downarrow}(\bm{k},1-m,1-n)
      ,\,\label{Fdefdown}
    \end{align}
  \end{subequations}
  and rewrite Eq.~(\ref{kraumbes}) as
  \begin{align}
    n_{\bm{k}\sigma}(g,n,m)
    &=
    n_{\bm{k}\sigma}^{0}(n,m)
    +
    \frac{1}{(1+g)^2}
    \sum_{p=0}^{\infty}(g^2-1)^{p+2}\,F_{p\sigma}(\bm{k},n,m)
    .\,\label{nkumsummiert}
  \end{align}
  This expression appears on the left-hand sides of
  Eqs.~(\ref{kuprelation}) and (\ref{kdownrelation}), while their
  right-hand sides take the form
  \begin{subequations}\label{nktransformiert}
    \begin{align}
      1-n_{\bm{Q}-\bm{k}\uparrow}(g^{-1},1-m,1-n)
      &=
      n_{\bm{k}\uparrow}^{0}(n,m)
      +
      \frac{1}{(1+g)^2}
      \sum_{p=0}^{\infty}
      \frac{(g^2-1)^{p+2}}{(-g^2)^{p+1}}
      \,
      \bar{F}_{p\uparrow}(\bm{k},n,m)
      ,\,\label{nkuptransformiert}
      \\
      n_{\bm{k}\downarrow}(g^{-1},1-m,1-n)
      &=
      n_{\bm{k}\downarrow}^{0}(n,m)
      -
      \frac{1}{(1+g)^2}
      \sum_{p=0}^{\infty}
      \frac{(g^2-1)^{p+2}}{(-g^2)^{p+1}}
      \,
      \bar{F}_{p\downarrow}(\bm{k},n,m)
      .\,\label{nkdowntransformiert}
    \end{align}
  \end{subequations}
  Expanding $(g^2)^{-p-1}$ in powers of $(g^2-1)$
  {[Eq.~(\ref{ginvexpansion})]}, comparing coefficients, and combining
  both cases, we find
  \begin{align}
    F_{p\sigma}(\bm{k},n,m)
    &=
    -\,\text{sgn}(\sigma)(-1)^{p}
    \sum_{r=0}^{p}
    \binom{p}{r}
    \bar{F}_{r\sigma}(\bm{k},n,m)
    .\,\label{Frelation}
  \end{align}
  For clarity we will {}from now on label $f_{p\sigma}$,
  $h_{p\sigma}$, and $n_{\bm{k}\sigma}$ with the subscripts $\inside$
  and $\outside$, depending on whether the momentum lies inside or
  outside of the Fermi surface. We first simplify the equations for
  $f_{p\uparrow}$.  Using Eq.~(\ref{Frelation}) together with
  (\ref{crelation}) we obtain
  \begin{align}
    h_{p+1\uparrow}(\bm{k},n,m)
    &=
    (-1)^{p}\sum_{r=1}^{p}
    \binom{p}{r}
    h_{r+1\uparrow}(\bm{Q}-\bm{k},1-m,1-n)
    ,\,\label{huprelation}
  \end{align}
  valid for all $\bm{k}$, relating $h_{p\uparrow}^{\inside}$ and
  $h_{p\uparrow}^{\outside}$.  For the minority spin a similar
  calculation yields
  \begin{subequations}
    \label{fdownrelation}
    \begin{align}
      f_{p\downarrow}^{\inside}(\bm{k},n,m)
      &=
      (-1)^{p}
      \sum_{r=0}^{p}
      \binom{p}{r}
      f_{r\downarrow}^{\inside}(\bm{k},1-m,1-n)
      ,\,\label{fdownrelin}
      \\
      f_{p+2\downarrow}^{\outside}(\bm{k},n,m)
      &=
      (-1)^{p}\sum_{r=0}^{p}
      \binom{p}{r}
      f_{r+2\downarrow}^{\outside}(\bm{k},1-m,1-n)
      ,\,\label{fdownrelout}
    \end{align}
  \end{subequations}
  i.~e., there is a relation between the $f_{p\downarrow}$ for momenta
  inside of the Fermi surface, and another relation for momenta
  outside of the Fermi surface, each linking the cases $n+m\leq1$ and
  $n+m\geq1$.  Note that the relations in
  Eqs.~(\ref{huprelation})-(\ref{fdownrelation}) are valid in
  arbitrary dimensions for lattices with hopping between $A$ and $B$
  sublattices only.

  \section{Analytic evaluation in $D=1$}
  \label{sec:evaluation}
  
  In the remainder of the paper we consider the GWF for the Hubbard
  chain with a symmetric dispersion ${\epsilon}_{k}={\epsilon}_{-k}$
  that increases monotonically with $|k|$. This implies that
  $n_{k\sigma}=n_{-k\sigma}$; therefore we only consider $k\geq0$.
  For convenience we assume that the zero of energy is chosen such
  that the center of the band,
  $\bar{\epsilon}\defeq2\int_{0}^{1/2}\!dk\,\epsilon_{k}$, is zero.
  
  For such dispersions the free Fermi sea described by the starting
  wave function $|\Phi_{0}\rangle$ is centered around $k=0$ and is
  simply connected, i.~e.,
  $n_{k\sigma}^{0}=\Theta(k_{\text{F}\sigma}-|k|)$ is a step function.
  (We follow the convention of MV of measuring $k$ in units of
  $2\pi/a$, where $a$ is the lattice spacing; the first Brillouin zone
  is the interval $[-1/2;1/2]$, the reciprocal lattice vectors $K$ are
  integers, and the nesting vector $Q=1/2$.)  For this Fermi surface
  topology the Fermi momentum,
  $k_{\text{F}\sigma}=n_{\sigma}/2=(n+\text{sgn}(\sigma)m)/4$, only
  depends on the particle density $n_{\sigma}$, i.~e., the particular
  form of the dispersion is irrelevant. In general this simplification
  occurs only in dimension $D=1$; higher-dimensional tight-binding
  dispersions usually do not depend only on $|\bm{k}|$, although this
  symmetry can be artificially imposed to allow the construction of a
  dispersion {}from a given density of states.\cite{Kollar01a}
  
  Since the values of the diagrams $c_{p}$ and $f_{p\sigma}$ are
  completely determined by the function $n_{k\sigma}^{0}$ (which is
  independent of the dispersion as long as $\epsilon_{k}$ is
  increasing with $|k|$) the relations derived in
  Sec.~\ref{subsec:trafos}, and hence the relations in
  Eqs.~(\ref{erstecrelation}), (\ref{crelation}), (\ref{huprelation}),
  (\ref{fdownrelation}), are valid for $\emph{all}$ increasing and
  symmetric dispersions in $D=1$.
  
  The following analytic calculation of GWF expectation values with
  magnetization $m\not=0$ is based on the corresponding calculation
  for $m$ $=$ $0$ by MV. The calculation for $m$ $=$ $0$ was made
  possible by exploiting the relations following {}from canonical
  transformations, the polynomial form of the diagrams and their
  continuity as functions of $k$ and $n$, and an analysis of the
  contribution of Umklapp processes. We will now use very similar
  methods to express the double occupation $d(g,n,m)$ and the momentum
  distribution $n_{k\sigma}(g,n,m)$ in terms of the known quantities
  for $m$ $=$ $0$.
  
  In Section A we review the results of MV and present new closed
  formulas for $n_{k\sigma}$ for zero magnetization. For the magnetic
  case the double occupation, the momentum distribution, and its
  discontinuity at the Fermi surface are calculated in Sections B, C,
  and D. The variational energy is evaluated in Section E.

  \subsection{Zero magnetization}\label{subsec:nonmagn}
  
  For the non-magnetic case the diagrams $c_{p}$ and $h_{p\sigma}$
  were already calculated by MV. For $m$ $=$ $0$, $n\leq1$, $0\leq
  k\leq1/2$, the results may be summarized as
  \begin{align}
    c_{p}(n,0)
    &=
    \gamma_{p}\,n^{p+1}
    ,\,\label{cnonmagn}
    \\
    f_{p\sigma}(k,n,0)
    &=
    h_{p\sigma}(k,n,0)+c_{p-1}(n,0)
    \\
    &=
    \begin{cases}
      n^{p}\,R_{p}({\textstyle{\frac{k}{n}}})
      &\text{for }
      k<\frac{n}{4}
      \\
      n^{p}{[Q_{p}({\textstyle{\frac{k}{n}}})+\gamma_{p-1}]}
      &\text{for }
      \frac{n}{4}<k\leq\text{min}({\textstyle{\frac{3n}{4}}},
      1\!-\!{\textstyle{\frac{3n}{4}}})
      \\
      n^{p}\,\gamma_{p-1}
      &\text{for }
      \frac{3n}{4}\leq k\leq\frac{1}{2}
      \\
      n^{p}{[Q_{p}({\textstyle{\frac{k}{n}}})
        +Q_{p}({\textstyle{\frac{1-k}{n}}})+\gamma_{p-1}]}
      &
      \text{for }
      1\!-\!\frac{3n}{4}\leq k\leq\frac{1}{2}
    \end{cases}
    \!,\;\;p\geq 2
    .\,\label{fnonmagn}
  \end{align}
  Here and below $\gamma_{p}=(-1)^{p+1}/{[2(p+1)]}$, and $R_{p}(k)$
  and $Q_{p}(k)$ are certain polynomials of degree $p$ in $k$. Note
  that at $k_{\text{F}}=n/4$ both $f_{p\sigma}$ and $h_{p\sigma}$ are
  discontinuous, and $h_{p\sigma}=0$ for $k\geq3n/4$. The contribution
  with momentum $1-k$ in Eq.~(\ref{fnonmagn}) is due to Umklapp
  processes.  {}From the diagrammatic series in
  Eq.~(\ref{entwicklungen}) MV's result for the double occupation is
  \begin{align}
    d(g,n,0)
    =
    \frac{g^2}{2(1-g^2)^2}
    \left(
      -\ln\!{[1-(1-g^2)n]}
      -(1-g^2)n
    \right)
    \!,\,\label{doccnonmagn}
  \end{align}
  and the momentum distribution $n_{k\sigma}$ is given by
  \begin{align}
    n_{k\sigma}^{\inside}(g,n,0)
    &=
    1
    -
    \frac{1-g}{1+g}\;\frac{n}{2}
    +
    \frac{g^2}{(1+g)^2}
    \left[
      {\cal R}_{0}(\textstyle\frac{4k}{n},(1-g^2)n)-1
    \right]
    \!,\,\label{nonmagninresult}
    \\
    n_{k\sigma}^{\outside}(g,n,0)
    &=
    -\,
    \frac{1-g}{1+g}\;\frac{n}{2}
    +
    \frac{1}{(1+g)^2}
    \big[
      -\textstyle\frac{1}{2}\ln\!{[1-(1-g^2)n]}
      \nonumber\\&~~~~~
      +
      {\cal Q}_{0}(\textstyle\frac{4k-2n}{n},(1-g^2)n)
      +
      {\cal Q}_{0}(\textstyle\frac{4(1-k)-2n}{n},(1-g^2)n)
      \big]
    ,\,\label{nonmagnoutresult}
  \end{align}
  for $k<k_{\text{F}}$ and $k>k_{\text{F}}$, respectively. Here we
  introduced the generating functions
  \begin{align}
    {\cal R}_{0}(x,z)
    &\defeq
    \sum_{p=0}^{\infty}
    R_{p}({\textstyle\frac{x}{4}})\,(-z)^{p}
    ,\,\label{rgendef}
    \\
    {\cal Q}_{0}(x,z)
    &\defeq
    \sum_{p=0}^{\infty}
    Q_{p}({\textstyle\frac{2+x}{4}})\,(-z)^{p}
    ,\,\label{qgendef}
  \end{align}
  with the convention that ${\cal R}_{0}(x,z)$ and
  ${\cal Q}_{0}(x,z)$ are zero for $|x|>1$.
  
  In MV the coefficients of the polynomials $R_{p}(k)$ and $Q_{p}(k)$
  had to be calculated recursively to obtain the momentum
  distribution, and convergence was problematic for $(1-g^2)n$ close
  to 1. In Appendix~\ref{app:polynomials} we show how the recursion
  equations for $R_{p}(k)$ can in fact be solved in closed form
  {[Eqs.~(\ref{r-explicit1})-(\ref{r-explicit4})]}, while $Q_{p}(k)$
  can be expressed in terms of them and their integrals. The
  generating functions are calculated as
  \begin{align}
    {\cal R}_{0}(x,z)
    &=
    \frac{4/\pi}{\sqrt{(2-z)^2-(xz)^2}}
    \,
    \bm{K}
    \Big(\frac{z\sqrt{1-x^2}}{\sqrt{(2-z)^2-(x z)^2}}\Big)
    ,\,\label{r-function}
    \\
    {\cal Q}_{0}(x,z)
    &=
    {\cal W}_{0}(x,z)
    +\,
    \frac{z}{2}
    \left[
      (1-x)\,{\cal R}_{0}(x,z)+{\cal R}_{1}(x,z)
    \right]
    \!,\,\label{q-function}
  \end{align}
  for $|x|\leq1$. Here $\bm{K}(k)$ $=$
  $\int_{0}^{\pi/2}(1-k^2\sin^2\phi)^{-1/2}d\phi$
  is the complete elliptic integral of the first kind,
  ${\cal W}_{0}(x,z)$ is an auxiliary function,
  \begin{align}
    {\cal W}_{0}(x,z)
    &\defeq
    \frac{x-1}{2}\,{\cal R}_{0}(x,z)
    +
    \frac{z-2}{4}\,{\cal R}_{1}(x,z)
    +
    \frac{z(z-1)}{2}\,
    \dot{\cal R}_{1}(x,z)
    ,\,\label{w-function}
  \end{align}
  where the dot indicates a partial derivative with respect to second
  argument, and ${\cal R}_{j}(x,z)$ is the repeated integral of ${\cal
    R}_{0}(x,z)$, defined by ($j\geq0$)
  \begin{align}
    {\cal R}_{j+1}(x,z)
    &\defeq
    \int_{1}^{x}\!dx^{\prime}
    \;{\cal R}_{j}(x^{\prime},z)
    =
    \frac{1}{j!}
    \int_{1}^{x}\!dx^{\prime}
    \,(x-x')^{j}
    \,{\cal R}_{0}(x^{\prime},z)
    .\,\label{r-integral}
  \end{align}
  Below we will also need the following integral,
  \begin{align}
    {\cal Q}_{1}(x,z)
    &\defeq
    \int_{1}^{x}\!dx^{\prime}
    \;{\cal Q}_{0}(x^{\prime},z)
    \nonumber\\&
    =
    \frac{(1-z)(x-1)}{2}\,{\cal R}_{1}(x,z)
    +\frac{5z-4}{4}\,{\cal R}_{2}(x,z)
    +\frac{z(z-1)}{2}\,{\frac{\partial}{\partial z}}\,{\cal R}_{2}(x,z)
    .\,\label{q-integral}
  \end{align}
  In Appendix~\ref{app:polynomials} we provide an explicit expression
  for ${\cal R}_{j}(x,z)$
  {[Eqs.~(\ref{r-integral-alt})-(\ref{r-integral-alt-polynomial})]},
  as well as other relations. Here we note in particular the following
  functional relations, which are obtained {}from
  Eqs.~(\ref{r-function}) and (\ref{q-function}),
  \begin{align}
    {\cal R}_{j}(x,\textstyle\frac{z}{z-1})
    &=
    (1-z)\,{\cal R}_{j}(x,z)
    ,\,\label{r-fortsetzung}
    \\
    {\cal Q}_{j}(x,\textstyle\frac{z}{z-1})
    &=
    {\cal W}_{j}(x,z)
    .\,\label{q-fortsetzung}
  \end{align}
  These expressions analytically continue ${\cal R}_{j}(x,z)$ and
  ${\cal Q}_{j}(x,z)$ to $|z|>1$.

  \subsection{Non-zero magnetization: double occupation}
  \label{subsec:magndocc}
  
\begin{figure}
    \centerline{\includegraphics[height=10cm]{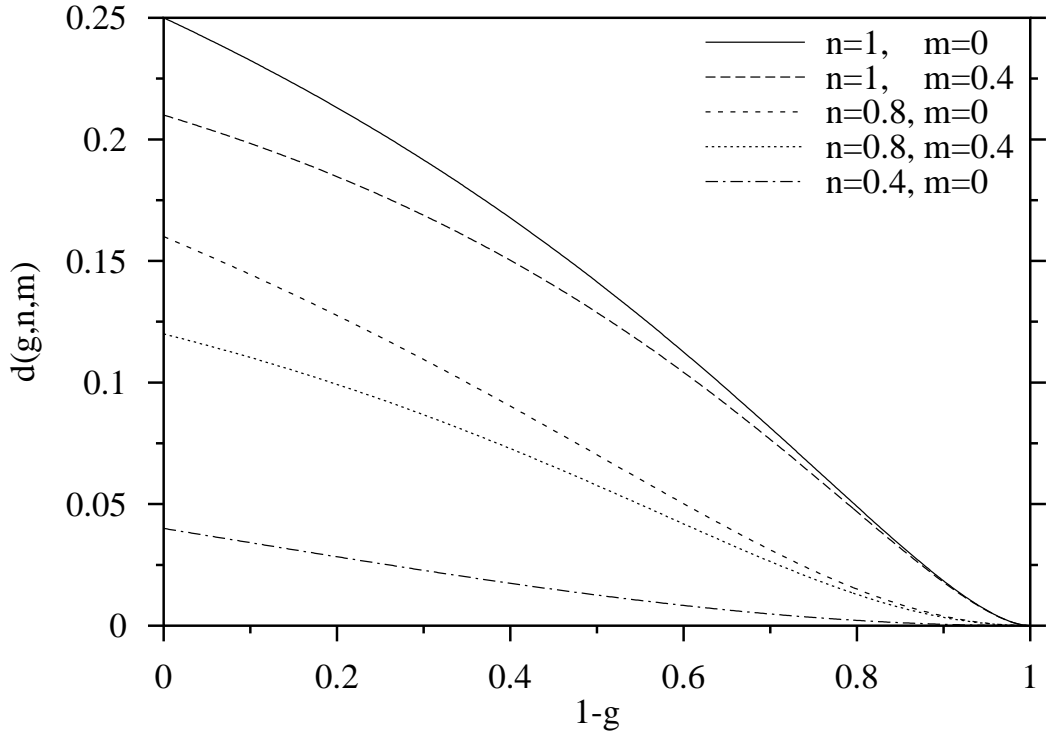}}
    \caption{Double occupation $d(g,n,m)$ as a function of
      the variational parameter $g$ for various densities $n$ and
      magnetizations $m$.  Note that
      $d(g,n,m)=d(g,n,0)-d(g,m,0)$.\label{fig:docc}}
  \end{figure}

  To calculate the double occupation one needs the diagrams $c_{p}$.
  Using the methods of MV one can show that for $m>0$ Umklapp
  processes occur in the graphs for $c_{p}(n,m)$ only if $n\geq1$.
  This implies that for $n<1$ $c_{p}(n,m)$ is a homogeneous function
  of $k_{\text{F}\uparrow}$ and $k_{\text{F}\downarrow}$ of order
  $p+1$, since every contributing graph contains $p+1$ free momenta.
  Since $k_{\text{F}\sigma}=(n+\text{sgn}(\sigma)m)/4$ it follows that
  $c_{p}(n,m)$ can be written as a linear combination of terms
  $n_{\uparrow}^{r}n_{\downarrow}^{p+1-r}$ with $0\leq r\leq p+1$,
  i.~e., $c_{p}(n,m)/n^{p+1}$ is a polynomial in $m/n$ of degree
  $p+1$.  This is also the case for $n$ $=$ $1$ since $d(g,n,m)$ is
  continuous at $n$ $=$ $1$ {[see Eq.~(\ref{bigdensdocc})]}. Therefore
  we can write the polynomial simply as
  $c_{p}(n,m)/n^{p+1}=c_{p}(1,m/n)$ for $n\leq1$.  For $n$ $=$ $1$,
  however, we can calculate $c_{p}(1,m)$ {}from Eq.~(\ref{crelation})
  and Eq.~(\ref {cnonmagn}):
  \begin{align}
    c_{p}(1,m)
    &=
    (-1)^{p}\sum_{r=0}^{p}
    \binom{p}{r}
    c_{r}(1-m,0)
    \nonumber\\
    &=
    \gamma_p\,(1-m^{p+1})
   .%
  \end{align}
  The polynomial structure of $c_{p}(n,m)$ then implies
  \begin{align}
    c_{p}(n,m)
    &=
    n^{p+1}c_{p}(1,{\textstyle{\frac{m}{n}}})
    =
    \gamma_p\,(n^{p+1}-m^{p+1})
    \nonumber\\
    &=
    c_{p}(n,0)-c_{p}(m,0)
    .%
  \end{align}
  Summation of the series in Eq.~(\ref{doppelbes}) yields the simple
  result
  \begin{align}
    d(g,n,m)=d(g,n,0)-d(g,m,0).
  \end{align}
  It is remarkable that the double occupation at density $n$ and
  magnetization $m$ is obtained as the difference between the double
  occupation without magnetization at density $n$ and \emph{density}
  $m$. With the closed form of $d(g,n,0)$ taken {}from
  Eq.~(\ref{doccnonmagn}) we finally obtain for the double occupation,
  valid for $0\leq m\leq n\leq1$,
  \begin{align}
    d(g,n,m)
    &=
    \frac{g^2}{2(1-g^2)^2}
    \left(
      \ln\frac{1-(1-g^2)m}{1-(1-g^2)n}
      -(1-g^2)(n-m)
    \right)
    \!,\,\label{doccergebnis}
  \end{align}
  The double occupation is shown in Fig.~\ref{fig:docc} for various
  parameter values. In the limit of strong correlation ($g\to0$) it
  behaves as
  \begin{subequations}
    \begin{align}
      d(g,n<1,m\leq n)
      &=
      \frac{g^2}{2}
      \left[
        \ln\frac{1-m}{1-n}
        -(n-m)
      \right]
      +O(g^4)
      ,
      \\
      d(g,n=1,m<1)
      &=
      g^2
      \left[
        \ln\frac{1}{g}
        +
        \frac{\ln(1-m)-(1-m)}{2}
        \right]
      +O(g^4\ln g)
      ,
    \end{align}
  \end{subequations}
  i.~e., the double occupation is \emph{non-analytic} in the limit
  $n\to1$, $g\to0$, as in the non-magnetic case (see MV).

  \subsection{Non-zero magnetization: momentum distribution}
  \label{subsec:magnkocc}
  
  To calculate the momentum distribution $n_{k\sigma}$ one needs the
  diagrams $f_{p\sigma}$ or $h_{p\sigma}$.  In the following
  subsections we distinguish the cases of majority and minority spin,
  as well as whether $k$ lies inside or outside of the Fermi surface.
  
  For finite magnetization, $m>0$, we make the following observations,
  similar in spirit to those of MV for the non-magnetic case. As a
  function of $k$, $f_{p\sigma}(k,n,m)$ and $h_{p\sigma}(k,n,m)$ are
  discontinuous at $k=k_{\text{F}\sigma}$, since the one-particle
  irreducible graphs contain a factor $n_{k\sigma}^{0}$. For $k\geq
  k_{\text{F}\sigma}+2k_{\text{F}\bar{\sigma}}$ momentum conservation
  at the outer vertices of $h_{p\sigma}$ cannot be fulfilled, so that
  in this case $h_{p\sigma}(k,n,m)=0$. For $k\geq 1-k_{\text{F}\sigma
    }-2k_{\text{F}\bar{\sigma}}\;(\geq k_{\text{F}\sigma})$ Umklapp
  processes occur and yield an additional contribution to
  $h_{p\sigma}$ of normal processes with external momentum $1-k$. In
  the absence of Umklapp processes $f_{p\sigma}$ and $h_{p\sigma}$ are
  homogeneous functions of $k_{\text{F}\uparrow}$ and
  $k_{\text{F}\downarrow}$ of order $p$, since every contributing
  graph contains $p$ free momenta. Then $f_{p\sigma}/n^{p}$ and
  $h_{p\sigma}/n^{p}$ are polynomials in $k/n$ and $m/n$ of degree
  $p$, and due to momentum conservation at outer vertices different
  polynomials occur depending on whether $k$ is larger or smaller than
  $\pm(2k_{\text{F}\bar{\sigma}}-k_{\text{F}\sigma})$. Furthermore,
  different polynomials for $h_{p\downarrow}^{\outside}/n^{p}$ occur
  also depending on how $k$ compares to $3k_{\text{F}\downarrow}$ (see
  subsection~4 below and
  Appendix~\ref{app:down-out-sector-structure}).

  \subsubsection{Majority spins inside of the Fermi surface:
    $n_{k\uparrow}^{\inside}(g,n,m)$}
  \label{subsubsec:up-in}
  
  For $0\leq k<k_{\text{F}\uparrow}=(n+m)/4$ and $n$ $=$ $1$ the
  diagrams $h_{p\uparrow}^{\inside}$ can be obtained {}from
  Eq.~(\ref{huprelation}) in terms of the known functions
  $h_{p\uparrow}^{\outside}$ of the non-magnetic case
  [Eq.~(\ref{fnonmagn})]:
  \begin{align}
    h_{p+1\uparrow}^{{\inside}}(k,1,m)
    &=
    (-1)^{p}\sum_{r=1}^{p}
    \binom{p}{r}
    h_{r+1\uparrow}^{\outside}({\textstyle{\frac{1}{2}}}-k,1-m,0)
    ,\,\label{hupinrelation}
    \\
    &=
    (-1)^{p}\sum_{r=1}^{p}
    \binom{p}{r}
    (1-m)^{r+1}
    \nonumber\\&~~~~~
    \times
    \begin{cases}
      Q_{r+1}(\frac{\frac{1}{2}-k}{1-m})
      & \text{for }
      \frac{|3m-1|}{4}\leq k<\frac{1+m}{4}
      \\
      0
      & \text{for }
      0\leq k\leq \frac{3m-1}{4}
      \\
      {[Q_{r+1}(\frac{\frac{1}{2}-k}{1-m})
        +Q_{r+1}(\frac{\frac{1}{2}+k}{1-m})]}
      & \text{for }
      0\leq k\leq \frac{1-3m}{4}
    \end{cases}
    \!.\,\label{hupingleichung}
  \end{align}
  Since Umklapp processes do not occur for $k<k_{\text{F}\uparrow}$,
  $h_{p\uparrow}^{\inside}(k,n,m)/n^{p}$ is a polynomial in $k/n$ and
  $m/n$. Hence we can obtain $h_{p+1\uparrow}^{\inside}/n^{p+1}$ by
  replacing $k$ by $k/n$ and $m$ by $m/n$ in
  Eq.~(\ref{hupingleichung}). This yields
  \begin{align}
    h_{p+1\uparrow}^{\inside}(k,n,m)
    &=
    (-1)^{p}\sum_{r=1}^{p}
    \binom{p}{r}
    (n-m)^{r+1}n^{p-r}
    \nonumber\\&~~~~~
    \times
    \begin{cases}
      Q_{r+1}(\frac{\frac{n}{2}-k}{n-m})
      & \text{for }
      \frac{|3m-n|}{4}\leq k<\frac{n+m}{4}
      \\
      0
      & \text{for }
      0\leq k\leq \frac{3m-n}{4}
      \\
      {[Q_{r+1}(\frac{\frac{n}{2}-k}{n-m})
        +Q_{r+1}(\frac{\frac{n}{2}+k}{n-m})]}
      & \text{for }
      0\leq k\leq \frac{n-3m}{4}
    \end{cases}
    \!.\,\label{hupinergebnis}
  \end{align}
  Note that $h_{p+1\uparrow}^{\inside}(k,n,m)$ is continuous for all
  $k<k_{\text{F}\uparrow}$ due to $Q_{p}(3/4)=0$ (see MV). The normal
  processes contribute differently depending on how $k$ compares to
  $\pm(2k_{\text{F}\downarrow}-k_{\text{F}\uparrow})$, as expected.

  Since $h_{p\uparrow}^{\inside}$ is available for arbitrary orders of
  $p$, the series in Eq.~(\ref{kraumbes}) can now be summed, with the
  result, valid for $0\leq m<n\leq1$, $0\leq k<(n+m)/4$,
  \begin{align}
    n_{k\uparrow}^{\inside}(g,n,m)
    &=
    1-(1-g)^2
    \Big[
      \frac{n-m}{2}-d(g,n,m)
    \Big]
    \nonumber\\&~~~~~
    -\frac{g^2}{(1+g)^2}
    \Big[
      {\cal W}_{0}\!\left(
        \textstyle\frac{2m-4k}{n-m},\frac{(1-g^2)(n-m)}{1-(1-g^2)m}
      \right)
      +
      {\cal W}_{0}\!\left(
        \textstyle\frac{2m+4k}{n-m},\frac{(1-g^2)(n-m)}{1-(1-g^2)m}
      \right)
    \Big]
    ,\,\label{upinresult}
  \end{align}
  where the functional relation in Eq.~(\ref{q-fortsetzung}) was used.
  For $m$ $=$ $0$ this result reduces to Eq.~(\ref{nonmagnoutresult})
  by virtue of Eq.~(\ref{q-symmetric-part}).

  \subsubsection{Majority spins outside of the Fermi surface:
    $n_{k\uparrow }^{\outside}(g,n,m)$}
  \label{subsubsec:up-out}
  
  For $(n+m)/4<k\leq1/2$ we can deduce $h_{p\uparrow}^{\outside}$ at once
  by inserting $h_{p\uparrow}^{{\inside}}$ into
  Eq.~(\ref{huprelation}):
  \begin{align}
    h_{p+1\uparrow}^{\outside}(k,n,m)
    &=
    \sum_{r=1}^{p}
    \binom{p}{r}
    (n-m)^{r+1}(-m)^{p-r}
    \nonumber\\&~~~~~
    \times
    \begin{cases}
      Q_{r+1}(\frac{k-\frac{m}{2}}{n-m})
      & \text{for }
      \frac{n+m}{4}<k\leq\text{min}(\frac{3n-m}{4},
      1\!-\!\frac{3n-m}{4})
      \\
      0
      & \text{for }
      \frac{3n-m}{4}\leq k\leq \frac{1}{2}
      \\
      {[Q_{r+1}(\frac{k-\frac{m}{2}}{n-m})
        +Q_{r+1}(\frac{1-k-\frac{m}{2}}{n-m})]}
      & \text{for }
      1\!-\!\frac{3n-m}{4}\leq k\leq\frac{1}{2}
    \end{cases}
    \!,%
  \end{align}
  which is continuous for all $k>k_{\text{F}\uparrow}$.  In this
  sector Umklapp processes occur for
  $k\geq1-2k_{\text{F}\downarrow}-k_{\text{F}\uparrow}$ and contribute
  to $h_{p\uparrow}^{\inside}$ like normal processes with external
  momentum $1-k$.
  
\begin{figure}
    \centerline{\includegraphics[height=20cm]{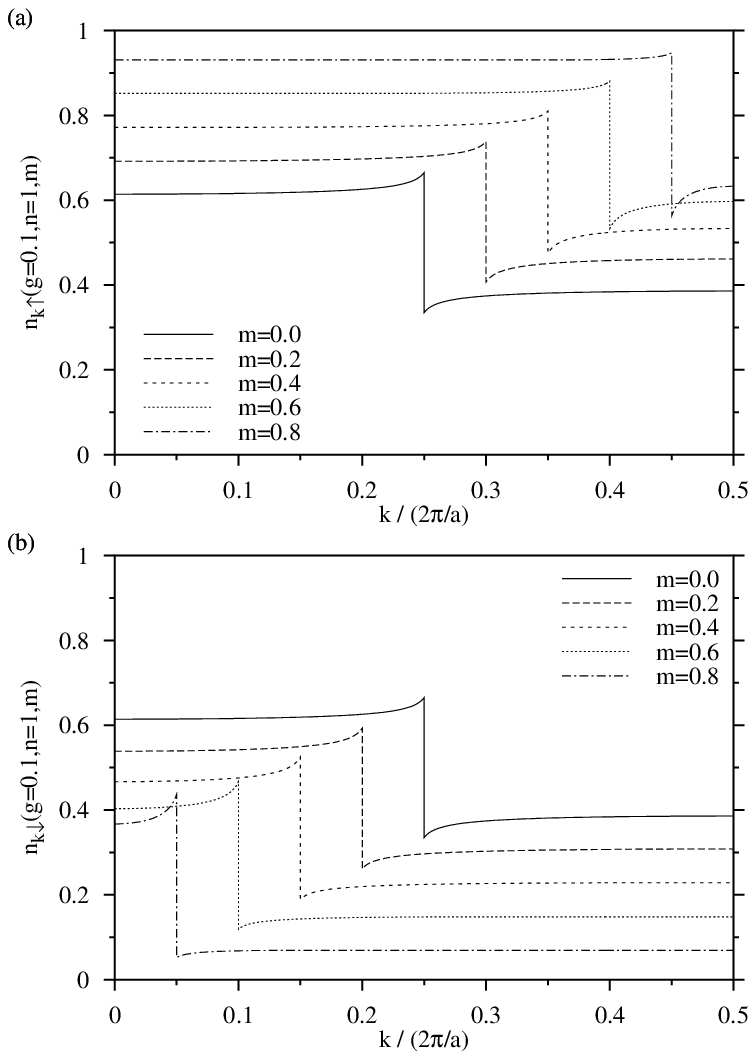}}
    \caption{Momentum distribution $n_{k\sigma}$ of (a) majority and
      (b) minority spin electrons, as a function of $k$ for $g$ $=$
      $0.1$ and density $n$ $=$ $1$.\label{fig:kocc10}}
  \end{figure}

\begin{figure}
    \centerline{\includegraphics[height=20cm]{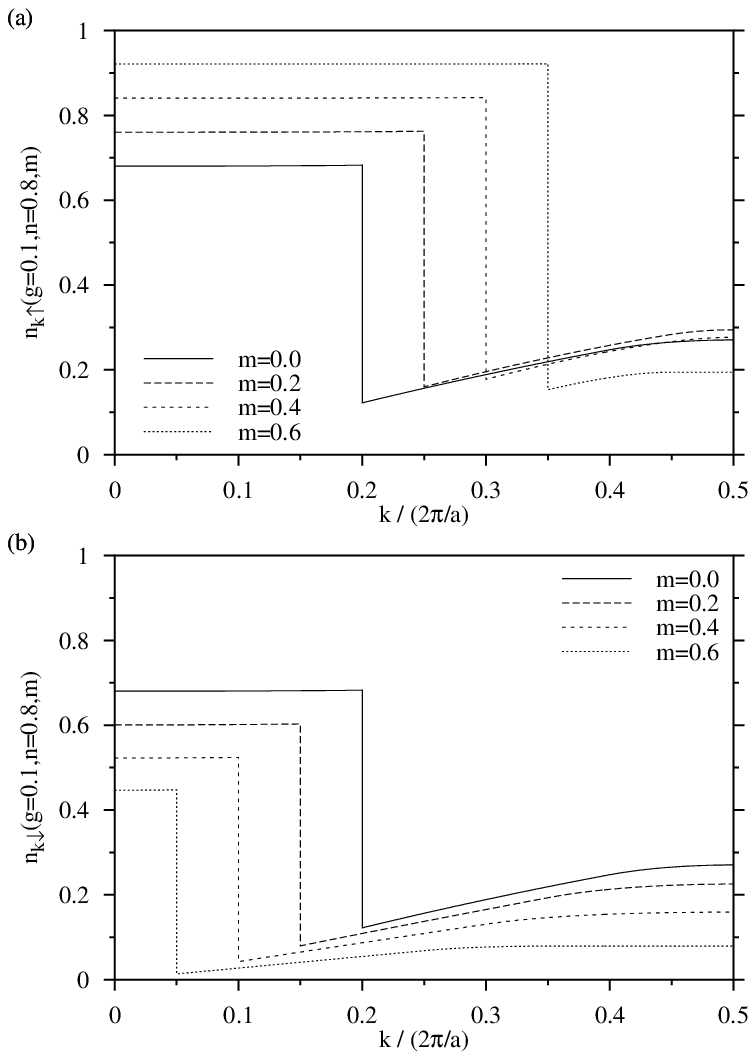}}
    \caption{Momentum distribution $n_{k\sigma}$ of (a) majority and
      (b) minority spin electrons as a function of $k$ for $g$ $=$
      $0.1$ and density $n$ $=$ $0.8$.\label{fig:kocc08}}
  \end{figure}

  Summing the series in Eq.~(\ref{kraumbes}), or applying
  Eq.~(\ref{kuprelation}) to Eq.~(\ref{upinresult}), yields the result,
  valid for $0\leq m<n\leq1$, $(n+m)/4<k\leq1/2$,
  \begin{align}
    n_{k\uparrow}^{\outside}(g,n,m)
    &=
    \frac{(1-g)^2}{g^2}\;d(g,n,m)
    \nonumber\\&~~~~~
    +\frac{1}{(1+g)^2}
    \left[
      {\cal Q}_{0}\!\left(
        \textstyle\frac{4k-2n}{n-m},
        \frac{(1-g^2)(n-m)}{1-(1-g^2)m}
      \right)
      +
      {\cal Q}_{0}\!\left(
        \textstyle\frac{4(1-k)-2n}{n-m},
        \frac{(1-g^2)(n-m)}{1-(1-g^2)m}
      \right)
    \right]
    \!,\,\label{upoutresult}
  \end{align}
  which alternatively can be derived directly {}from
  Eqs.~(\ref{kuprelation}), (\ref{upinresult}), and
  (\ref{q-fortsetzung}).  Note also that for $m$ $=$ $0$ this result
  reduces to Eq.~(\ref{nonmagnoutresult}). The momentum distribution
  $n_{k\uparrow}(g,n,m)$ has thus been determined in the entire
  parameter range. It is shown for $g$ $=$ $0.1$ and densities $n$ $=$
  1 and 0.8 in Fig.~\ref{fig:kocc10}a and \ref{fig:kocc08}a.

  \subsubsection{Minority spins inside of the Fermi surface:
    $n_{k\downarrow}^{\inside}(g,n,m)$}
  \label{subsubsec:down-in}
  
  For $0\leq k<k_{\text{F}\downarrow}=(n-m)/4$ and $n$ $=$ $1$ the
  diagrams $f_{p\downarrow}^{\inside}$ are given in terms of the known
  functions $f_{p\uparrow}^{\inside}$ of the non-magnetic case
  [Eq.~(\ref{fnonmagn})] according to Eq.~(\ref{fdownrelin}):
  \begin{align}
    f_{p\downarrow}^{\inside}(k,1,m)
    &=
    (-1)^{p}\sum_{r=0}^{p}
    \binom{p}{r}
    (1-m)^{r}R_{r}({\textstyle{\frac{k}{1-m}}})
    .\,\label{fdowningleichung}
  \end{align}
  Due to the absence of Umklapp processes for
  $k<k_{\text{F}\downarrow}$, $f_{p\downarrow}^{\inside}(k,n,m)/n^{p}$
  is a polynomial in $k/n$ and $m/n$. Therefore
  $f_{p\downarrow}^{\inside}/n^{p}$ is given by
  Eq.~(\ref{fdowningleichung}) with $k$ replaced by $k/n$ and $m$
  replaced by $m/n$. We then have
  \begin{align}
    f_{p\downarrow}^{\inside}(k,n,m)
    &=
    (-1)^{p}\sum_{r=0}^{p}
    \binom{p}{r}
    (n-m)^{r}n^{p-r}R_{r}({\textstyle{\frac{k}{n-m}}})
    ,\,\label{fdowninergebnis}
    \\
    &=
    \sum_{r=0}^{p}
    \binom{p}{r}
    (n-m)^{r}(-m)^{p-r}R_{r}({\textstyle{\frac{k}{n-m}}})
    ,\,\label{fdowninergebnis2}
  \end{align}
  where the second equation was obtained by inserting
  Eq.~(\ref{fdowninergebnis}) in Eq.~(\ref{fdownrelin}); it is
  equivalent to the functional relation in Eq.~(\ref{r-fortsetzung}).
  Note that for $m$ $=$ $0$ Eq.~(\ref{fdowninergebnis2}) reduces to
  (\ref{fnonmagn}).
  
  We use Eq.~(\ref{fdowninergebnis2}) in Eq.~(\ref{kraumbes}) to find
  the following expression for $n_{k\downarrow}^{\inside}$, valid for
  $0\leq m<n\leq1$, $0\leq k<(n-m)/4$,
  \begin{align}
    n_{k\downarrow}^{\inside}(g,n,m)
    &=
    1-\frac{1-g}{1+g}\;\frac{n+m}{2}
    +
    \frac{g^2}{(1+g)^2}
    \left[
      \frac{{\cal R}_{0}\!\left(
          \textstyle\frac{4k}{n-m},\frac{(1-g^2)(n-m)}{1-(1-g^2)m}
        \right)}{1-(1-g^2)m}
      -
      1
    \right]
    \!.\label{downinresult}
  \end{align}
  This expression reduces to Eq.~(\ref{nonmagninresult}) for $m$ $=$
  $0$.

  \subsubsection{Minority spins outside of the Fermi surface:
    $n_{k\downarrow}^{\outside}(g,n,m)$}
  \label{subsubsec:down-out}
  
  Finally we consider the case $(n-m)/4<k\leq1/2$, for which the
  calculation of $n_{k\downarrow}^{\outside}$ is somewhat more
  complicated. We begin with the special case $n$ $=$ $1$, for which
  the momentum distribution can be determined immediately {}from
  Eqs.~(\ref{kdownrelation}), (\ref{negmagnkocc}), and
  (\ref{kuprelation}), which combine to give
  \begin{align}
    n_{k\downarrow}^{\outside}(g,1,m)
    &=
    n_{k\downarrow}^{\outside}(g^{-1},1-m,0)
    =
    n_{k\uparrow}^{\outside}(g^{-1},1-m,0)
    =
    1-n_{\frac{1}{2}-k\uparrow}^{\inside}(g,1,m)
    .\,\label{downoutrelation-n=1}
  \end{align}
  Inserting Eq.~(\ref{upinresult}) we arrive at
  \begin{align}
    n_{k\downarrow}^{\outside}(g,1,m)
    &=
    (1-g)^2
    \left(
      \frac{1-m}{2}-d(g,1,m)
    \right)
    \nonumber\\&~~~~~
    +\frac{g^2}{(1+g)^2}
    \left[
      {\cal W}_{0}\!\left(
        \textstyle\frac{2m+4k-2}{1-m},
        \frac{(1-g^2)(1-m)}{1-(1-g^2)m}
      \right)
      +
      {\cal W}_{0}\!\left(
        \textstyle\frac{2m-4k+2}{1-m},
        \frac{(1-g^2)(1-m)}{1-(1-g^2)m}
      \right)
    \right]
    \!.\,\label{downoutresult-n=1}
  \end{align}  
  Next we consider arbitrary density and magnetization, $0\leq
  m<n\leq1$.  We make use of the following relation, which follows
  {}from Eqs.~(\ref{fdownrelout}) and (\ref{erstecrelation}),
  \begin{align}
    h_{p+2\downarrow}^{\outside}(k,n,m)
    &=
    -2\,c_{p+1}(n,m)-c_{p}(n,m)
    +(-1)^{p}\sum_{r=0}^{p}
    \binom{p}{r}
    h_{r+2\downarrow}^{\outside}(k,1-m,1-n)
    .\,\label{hdownrelout}
  \end{align}
  The diagrams $h_{p\downarrow}^{\outside}$ that appear in this
  equation can be written in terms of the contribution of normal
  processes, $N_p$, as
  \begin{align}
    h_{p\downarrow}^{\outside}(k,n,m)
    &=
    \begin{cases}
      n^{p}\,N_{p}({\textstyle\frac{k}{n}},
      {\textstyle\frac{m}{n}})
      &\text{for }
      \frac{n-m}{4}<k\leq
      \text{min}({\textstyle{\frac{3n+m}{4}}},
      1\!-\!{\textstyle{\frac{3n+m}{4}}})
      \\
      0
      &\text{for }
      \frac{3n+m}{4}\leq k\leq\frac{1}{2}
      \\
      n^{p}{[N_{p}({\textstyle\frac{k}{n}},
        {\textstyle\frac{m}{n}})
        +N_{p}({\textstyle\frac{1-k}{n}},
        {\textstyle\frac{m}{n}})]}
      &\text{for }
      1\!-\!\frac{3n+m}{4}\leq k\leq\frac{1}{2}
    \end{cases}
    \!.\,\label{hdownout}
  \end{align}
  In Appendix~\ref{app:down-out-sector-structure} we show that
  depending on $k$ and $m$, the function $N_p(k,m)$ is given piecewise
  by four polynomials in $k$ and $m$ of order $p$. The explicit
  determination of these polynomials is quite involved; it is
  presented in Appendix~\ref{app:down-out-sector-polynomials}.  The
  final result for $n_{k\downarrow}^{\outside}$, valid for $0\leq
  m<n\leq1$, $(n-m)/4<k\leq1/2$, can be written as
  \begin{align}
    n_{k\downarrow}^{\outside}(g,n,m)
    &=
    \frac{(1-g)^2}{g^2}\,d(g,n,m)
    +\frac{1}{(1+g)^2}
    \left[
      N(g,k,n,m)+N(g,1-k,n,m)
    \right]
    \!,\,\label{downoutresult}
  \end{align}
  where $N(g,k,n,m)$ is given by
  \begin{align}
    N(g,k,n,m)
    &=
    \begin{cases}
      N^{(1)}(g,k,n,m)
      &\text{for }
      \frac{n-m}{4}
      < k \leq
      \text{min}({\textstyle{\frac{n+3m}{4}}},
      {\textstyle{\frac{3(n-m)}{4}}})
      \\
      N^{(2)}(g,k,n,m)
      &\text{for }
      m \leq {\textstyle{\frac{n}{3}}}
      \text{ and }
      {\textstyle{\frac{n+3m}{4}}}
      \leq k \leq
      {\textstyle{\frac{3(n-m)}{4}}}
      \\
      N^{(3)}(g,k,n,m)
      &\text{for }
      m \geq {\textstyle{\frac{n}{3}}}
      \text{ and }
      {\textstyle{\frac{3(n-m)}{4}}}
      \leq k \leq
      {\textstyle{\frac{n+3m}{4}}}
      \\
      N^{(4)}(g,k,n,m)
      &\text{for }
      \text{max}({\textstyle{\frac{n+3m}{4}}},
      {\textstyle{\frac{3(n-m)}{4}}})
      \leq k \leq
      \frac{3n+m}{4}
      \\
      0
      &\text{for }
      \frac{3n+m}{4}
      \leq k
    \end{cases}
    \!,\,\label{N-cases}
  \end{align}
  with
  \begin{subequations}\label{N-functions}
    \begin{align}
      N^{(1)}(g,k,n,m)
      &=
      \frac{(1-(1-g^2)m)\,(3n-3m-4k)}{2\,(n-m)}
      \,
      {\cal Q}_{0}\!\left(
        \textstyle
        \frac{4k-2n+2m}{n-m},
        \frac{(1-g^2)(n-m)}{1-(1-g^2)m}
      \right)
      \nonumber\\&~~~~~
      +
      \frac{(n-3m)\,(1-g^2)+2}{4}
      \,
      {\cal Q}_{1}\!\left(
        \textstyle
        \frac{4k-2n+2m}{n-m},
        \frac{(1-g^2)(n-m)}{1-(1-g^2)m}
      \right)
      \nonumber\\&~~~~~
      +
      \frac{(n-m)\,(1-g^2)\,(1-(1-g^2)n)}{2\,(1-(1-g^2)m)}
      \,
      \dot{\cal Q}_{1}\!\left(
        \textstyle
        \frac{4k-2n+2m}{n-m},
        \frac{(1-g^2)(n-m)}{1-(1-g^2)m}
      \right)
      \nonumber\\&~~~~~
      +
      N^{(3)}(g,k,n,m)
      ,\,\label{N1-function}
      \\
      N^{(4)}(g,k,n,m)
      &=
      \frac{(1-(1-g^2)m)\,(4k-n-3m)}{2\,(n-m)}
      \,
      {\cal Q}_{0}\!\left(
        \textstyle
        \frac{4k-2n-2m}{n-m},
        \frac{(1-g^2)(n-m)}{1-(1-g^2)m}
      \right)
      \nonumber\\&~~~~~
      -
      \frac{(n-3m)\,(1-g^2)+2}{4}
      \,
      {\cal Q}_{1}\!\left(
        \textstyle
        \frac{4k-2n-2m}{n-m},
        \frac{(1-g^2)(n-m)}{1-(1-g^2)m}
      \right)
      \nonumber\\&~~~~~
      -
      \frac{(n-m)\,(1-g^2)\,(1-(1-g^2)n)}{2\,(1-(1-g^2)m)}
      \,
      \dot{\cal Q}_{1}\!\left(
        \textstyle
        \frac{4k-2n-2m}{n-m},
        \frac{(1-g^2)(n-m)}{1-(1-g^2)m}
      \right)
      \!,\,\label{N4-function}
      \\
      N^{(3)}(g,k,n,m)
      &=
      \frac{(n-m)(1-g^2)}{2}
      -
      \frac{1-(1-g^2)k}{2}\,
      \ln\frac{1-(1-g^2)m}{1-(1-g^2)n}
      ,\,\label{N3-function}
      \\
      N^{(2)}(g,k,n,m)
      &=
      N^{(1)}(g,k,n,m)
      -
      N^{(3)}(g,k,n,m)
      +
      N^{(4)}(g,k,n,m)
      ,\,\label{N2-function}
    \end{align}
  \end{subequations}
  and the dot again denotes derivative with respect to second
  argument.  It can be checked that Eq.~(\ref{downoutresult}) indeed
  reduces to (\ref{downoutresult-n=1}) for $m$ $=$ $0$.  Thus the
  momentum distribution $n_{k\downarrow}(g,n,m)$ has been determined
  for all parameters.  It is shown for $g$ $=$ $0.1$ and densities $n$
  $=$ 1 and 0.8 in Fig.~\ref{fig:kocc10}b and \ref{fig:kocc08}b.
  
  The calculation of the correlated momentum distributions
  $n_{k\sigma}$ is now complete. We remark that they are continuous
  functions of $k$, except at $k_{\text{F}\sigma}$, and are also
  continuous in $n$ and $m$ for fixed $k$. It can be checked that they
  obey the sum rule in Eq.~(\ref{sumrule}).

  \subsection{Discontinuity of the momentum distribution at the Fermi
    surface}\label{subsec:qdisc}
  
\begin{figure}
    \centerline{\includegraphics[height=20cm]{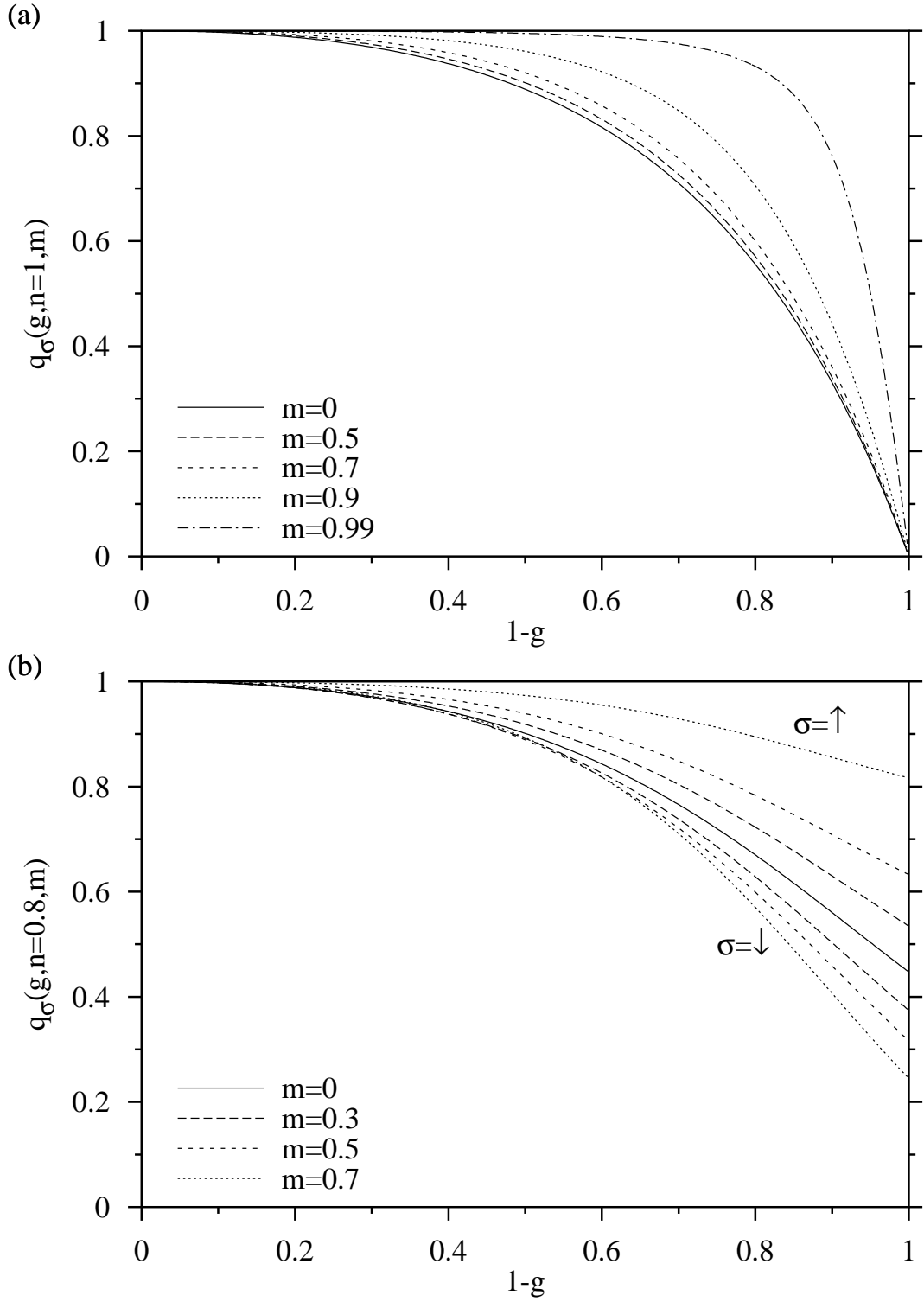}}
    \caption{Discontinuity $q_{\sigma}$ of the momentum 
      distribution at the Fermi vector for densities (a) $n$ $=$ $1$
      (in this case $q_\uparrow=q_\downarrow$) and (b) $n$ $=$
      $0.8$.\label{fig:qdisc}}
  \end{figure}

  It suffices to calculate the discontinuity $q_{\sigma}$ of
  $n_{k\sigma}$ at the Fermi surface for $0\leq m<n\leq1$ {[see
    Eq.~(\ref{qdisctrafo})]}.  {}From our previous results we obtain
  the left and right limit of the momentum distribution at the Fermi
  vector as
  \begin{subequations}
    \begin{align}
      n_{k_{\text{F}\sigma}^{-}\sigma}(g,n,m)
      &=
      1
      -\frac{1-g}{1+g}\,n_{\bar{\sigma}}
      +\frac{g^2\,(G_\sigma^{-1}-1)}{(1+g)^2}
      ,\,\label{nkf-minus}
      \\
      n_{k_{\text{F}\sigma}^{+}\sigma}(g,n,m)
      &=
      -\frac{1-g}{1+g}\,n_{\bar{\sigma}}
      +\frac{1-G_\sigma}{(1+g)^2}
      ,\,\label{nkf-plus}
    \end{align}
  \end{subequations}
  where the abbreviation $G_\sigma$ is defined as
  \begin{align}
    G_\sigma
    &\defeq
    \sqrt{
      [1-(1-g^2)n]
      [1-(1-g^2)m]^{-\text{sgn}(\sigma)}
      }
    .\,\label{Gdef}
  \end{align}
  For the discontinuity at the $\sigma$-spin Fermi surface we thus
  obtain
  \begin{align}
    q_{\sigma}(g,n,m)
    &=
    \frac{(g+G_\sigma)^2}{(1+g)^2\,G_\sigma}
    .\,\label{qdiscresult}
  \end{align}
  It follows that $q_{\sigma}$ vanishes only for a half-filled band
  without double occupation ($n$ $=$ $1$ and $g$ $=$ $0$); in this
  case there is exactly one particle at each site so that
  $n_{k\sigma}=1/2$ for all $k$. Note also that
  $q_\uparrow=q_\downarrow$ if $n$ $=$ $1$ (or, trivially, if $m$ $=$
  $0$ or $g$ $=$ $1$).  We plot $q_\sigma$ for $n$ $=$ $1$ and $n$ $=$
  $0.8$ in Fig.~\ref{fig:qdisc}.
  
  \subsection{Energy expectation value}\label{subsec:energy}  
  
  For any symmetric dispersion $\epsilon_{k}$, monotonically
  increasing with $|k|$, we can now calculate the energy expectation
  value per site, $E_{\text{G}}$ $=$
  $\langle\hat{H}\rangle_{\text{G}}/L$, of the one-dimensional Hubbard
  Hamiltionian (\ref{hamiltonian}), which is then minimized w.r.t.~$g$
  to find the optimal variational energy, $E_{\text{G}}^{\optimal}$,
  \begin{align}
    E_{\text{G}}^{\optimal}(n,m,U)
    &=
    \min_{0\leq g\leq1}
    E_{\text{G}}(g,n,m,U)
    ,\,\label{optimalenergy}
    \\
    E_{\text{G}}(g,n,m,U)
    &=
    2\int\limits_{0}^{1/2}\!dk\;
    \epsilon_k\,
    \sum_{\sigma}
    n_{k\sigma}(g,n,m)\,
    +\,
    U\,d(g,n,m)
    .\,\label{totalenergy}
  \end{align}
  Note that it follows {}from Eq.~(\ref{bigdenskocc}) that the total
  kinetic energy for dispersion $\epsilon_{k}$ at density $n>1$ can be
  calculated {}from the dispersion $-\epsilon_{1/2-k}$ at density
  $2-n$ using the formulas for $n_{k\sigma}$ and $d$ that are valid
  below half-filling.  
  
\begin{figure}
    \centerline{\includegraphics[height=20cm]{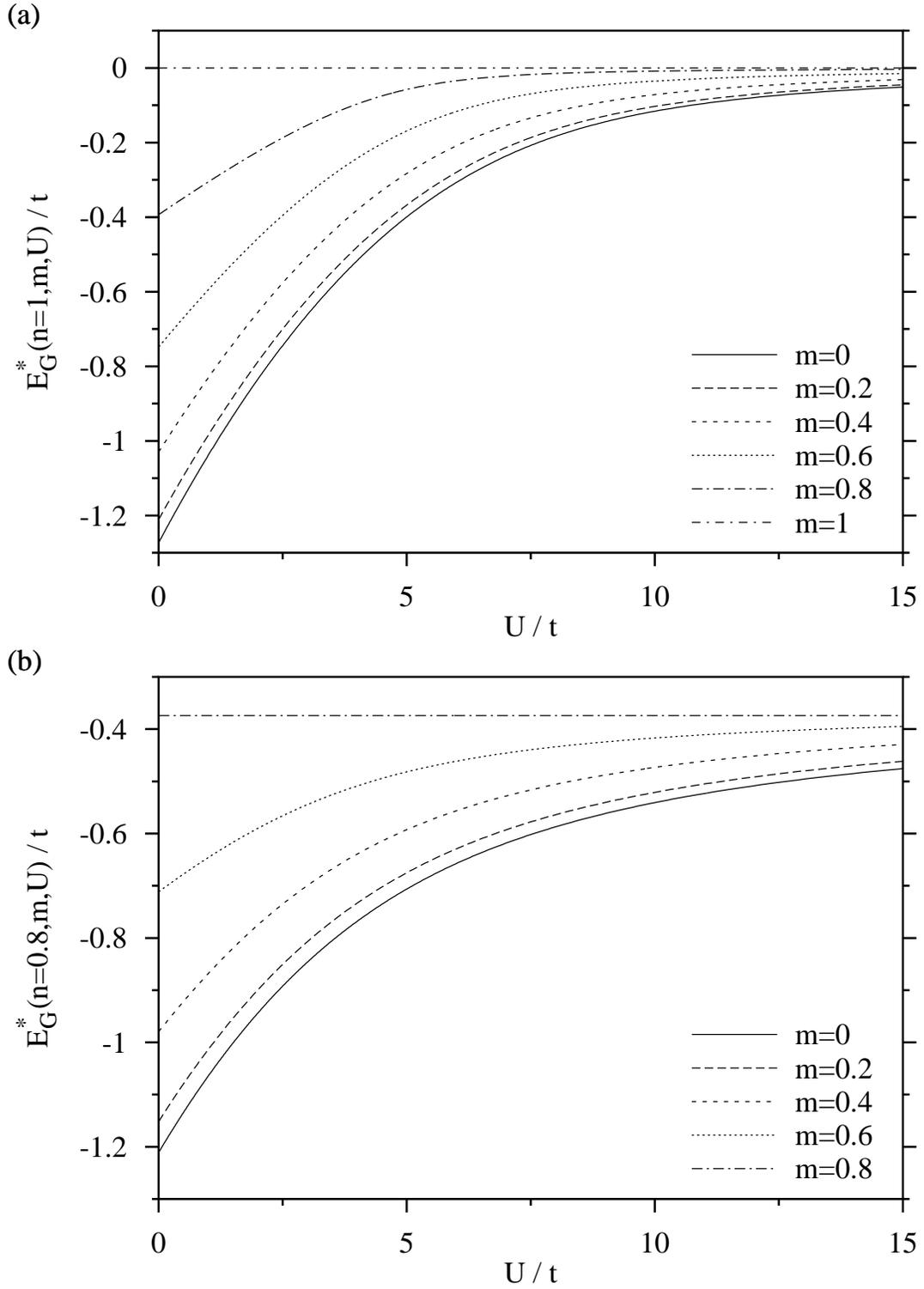}}
    \caption{Energy expectation value for
      the Hubbard chain with nearest-neighbor hopping $t>0$ for
      densities (a) $n$ $=$ $1$ and (b) $n$ $=$
      $0.8$.\label{fig:nn-energy-m}}
  \end{figure}

  For the Hubbard chain with nearest-neighbor hopping $t$ the
  dispersion relation in our notation is ${\epsilon}_{k} =
  -2t\cos(2\pi k)$. We assume $t$ $>$ $0$ without loss of generality,
  so that the dispersion is increasing with $|k|$ and our results for
  the Gutzwiller expectation values apply.  The optimal variational
  energy for this system is shown in Fig.~\ref{fig:nn-energy-m} for
  densities $n$ $=$ 1 and 0.8 for various magnetizations.  Note that
  at half-filling no Brinkman-Rice metal-insulator transition occurs
  at any finite $U$; i.~e., $g$ $=$ 0 is the optimal variational
  parameter only for $U$ $=$ $\infty$.  The variational result for the
  ground-state magnetization is determined in the next section.

  \section{Magnetic phase diagram of the Hubbard chain}
  \label{sec:results}
  
  In this section we determine the instability towards ferromagnetism
  for the Hubbard chain with nearest-neighbor hopping. Currently only
  homogeneous paramagnetic and ferromagnetic phases can be
  investigated analytically with the GWF in $D$ $=$ $1$; hence we do
  not consider antiferromagnetism or other broken symmetries.  We
  begin by examining the energy for the special cases of the
  paramagnetic state (i.~e., zero magnetization) and the fully
  polarized state.  The latter contains the minimum number of doubly
  occupied sites and is an eigenstate of $\hat{H}$, with eigenvalue
  \begin{align}
    E_{\text{FP}}(n,U)
    &=
    E_{\text{G}}(1,n,\min(n,2-n),U)
    \nonumber\\&
    =
    \begin{cases}
    \epsilon_0(n)
    &\text{for }0\leq n\leq1
    \\
    \epsilon_0(n-1)
    +
    U\,(n-1)
    &\text{for }1\leq n\leq2
    \end{cases}
    ,\,\label{energy-fp}
  \end{align}
  where $\epsilon_{0}(n_{\sigma}) \defeq 2\int_{0}^{n_{\sigma}/2}\!dk
  \,\epsilon_{k}$ is the kinetic energy of one spin species for the
  uncorrelated state. For the case of nearest-neighbor hopping we have
  $\epsilon_0(n_{\sigma})$ $=$ $-2t\sin(\pi n_{\sigma})/\pi$, $t$ $>$
  $0$.
  
\begin{figure}
    \centerline{\includegraphics[height=20cm]{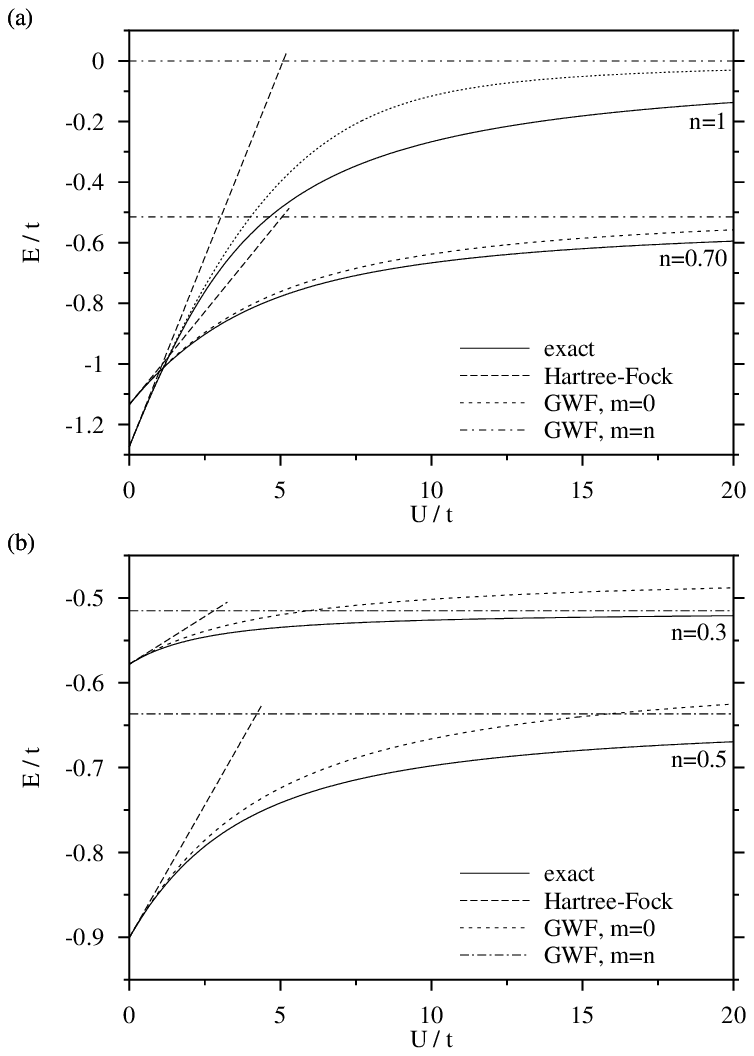}}
    \caption{Comparison of the total energy for
      the Hubbard chain with nearest-neighbor hopping and for
      densities (a) $n$ $=$ $0.7$, $1$ and (b) $n$ $=$ $0.3$,
      $0.5$.\label{fig:nn-energy-0}}
  \end{figure}

  In Fig.~\ref{fig:nn-energy-0} the exact ground-state energy,
  $E(n,m=0,U)$, obtained {}from the Bethe ansatz
  solution,\cite{Lieb68a} is compared to the Gutzwiller energy for
  zero and maximal polarization, at various densities. We also show
  the energy of the variational Hartree-Fock theory,
  $E_{\text{HF}}(n,m=0,U)$; it is contained as a special case in the
  results for the GWF,
  \begin{align}
    E_{\text{HF}}(n,m,U)
    &=
    E_{\text{G}}(1,n,m,U)
    \nonumber\\&
    =
    \epsilon_0\Big(\frac{n+m}{2}\Big)
    +
    \epsilon_0\Big(\frac{n-m}{2}\Big)
    +\frac{U}{4}\,(n^2-m^2)
    .\,\label{energy-hf}
  \end{align}
  As expected, the $g$-optimized GWF significantly improves upon
  Hartree-Fock theory but overestimates $E$ at large
  $U$.\cite{Metzner87a+88a} Since all spin configurations are
  degenerate for $U=\infty$, the exact ground-state energy coincides
  with $E_{\text{FP}}$ in this case; therefore the Gutzwiller energy
  $E_{\text{G}}^{\star}$ necessarily crosses the value $E_{\text{FP}}$
  at some finite value of the interaction $U_{c}$ (except for $n=1$).
  The existence of a finite critical interaction $U_{c}$ above which
  the GWF predicts a ferromagnetic ground state is in contrast to the
  Lieb-Mattis theorem,\cite{Lieb62a} which states that $m=0$ for the
  exact ground state (i.~e., $U_c=\infty$).  The reason for this
  overestimation of the instability of the paramagnetic state lies in
  the simple structure of the GWF, which controls only \emph{local}
  correlations and cannot describe the special correlated behavior in
  $D=1$ microscopically.
  
  The preceeding discussion only compared the variational energies for
  zero and full polarization. {}From our new results for the
  ferromagnetic GWF we can also study the stability of partially
  polarized ferromagnetic states. We first consider Hartree-Fock
  theory. A simple calculation shows that it predicts a fully
  polarized ground state for $U\geq U_c^{\text{HF}}(n)$, where
  \begin{align}
    U_c^{\text{HF}}(n)
    &=
    \begin{cases}
      16\sin(\pi n/2)\,{[1-\cos(\pi n/2)]}/(\pi n^2)
      &\text{for }0\leq n\leq1
      \\
      U_c^{\text{HF}}(2-n)
      &\text{for }1\leq n\leq2
    \end{cases}
    .\,\label{ucrit-HF}
  \end{align}
  This critical interaction $U_c^{\text{HF}}$ is smaller than that
  derived {}from the Stoner criterion,
  $\tilde{U}_c^{\text{HF}}(n)=1/N(\epsilon_{n/4})=2\pi\sin(\pi n/2)$,
  where $N(\epsilon)$ is the density of states.  Note that
  $E_{\text{HF}}$ as a function of $m$ never develops a local minimum
  at $m\neq0$. On the other hand, a maximum at finite $m$ occurs for
  $U$ $>$ $2(1-\cos(\pi n))/\min(n,2-n)$, which leads to a global
  minimum at full polarization already for $U$ $\geq$
  $U_c^{\text{HF}}$. The Stoner criterion, which merely signals a
  negative curvature of $E_{\text{HF}}$ at $m=0$ and does not take
  into account a finite magnetization, is thus irrelevant for the
  Hubbard chain with nearest-neighbor hopping.
  
\begin{figure}
    \centerline{\includegraphics[height=10cm]{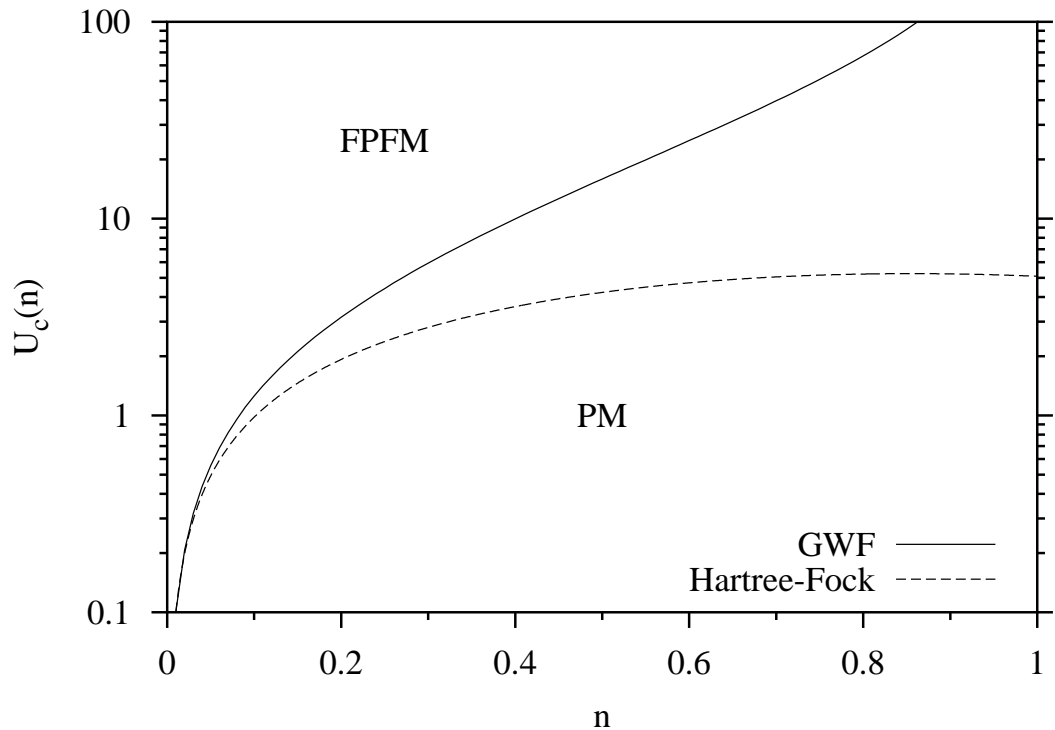}}
    \caption{Critical interaction $U_c(n)$ for the transition 
      {}from a paramagnetic to a fully polarized ferromagnetic state
      for the Hubbard chain with nearest-neighbor hopping as obtained
      {}from the Gutzwiller wave function. The Hartree-Fock result,
      $U_c^{\text{HF}}$ {[Eq.~(\ref{ucrit-HF})]}, is also shown; it
      has a shallow maximum at $n\approx0.856$. Note that $U_c(n)$ $=$
      $U_c(2-n)$.\label{fig:ucvsn}}
  \end{figure}

  For the Gutzwiller wave function we find that $E_{\text{G}}^{\star}$
  as a function of $m$ at fixed $U$ developments local extrema and
  global minima in a qualitatively similar fashion to $E_{\text{HF}}$.
  As a consequence the GWF also describes a discontinuous transition
  from the paramagnetic state to a state with full polarization at $U$
  $=$ $U_c(n)$. This critical interaction $U_c$ is shown in
  Fig.~\ref{fig:ucvsn}. Compared to Hartree-Fock theory we find
  agreement in the limit of small $n$. However, at intermediate
  densities the GWF predicts a significantly reduced ferromagnetic
  region.  In particular for $n$ $\to$ $1$ we have $U_c$ $\to$
  $\infty$, as expected from the previous discussion (see also
  Fig.~\ref{fig:nn-energy-0}a). Thus, in contrast to Hartree-Fock
  theory, the GWF does not exhibit a spurious ferromagnetic transition
  at half-filling, since it is able to avoid double occupation not
  only through a ferromagnetic polarization, but also by decreasing
  the variational parameter $g$.  Away from half-filling, however, the
  GWF predicts ferromagnetism for sufficiently large $U$, in contrast
  to the exact solution for the Hubbard chain.

  \section{Conclusion}
  \label{sec:conclusion}
  
  In this paper we presented general diagrammatic relations for the
  expectation values of the $D$-dimensional Hubbard model in terms of
  the Gutzwiller wave function (GWF) at \emph{non-zero} magnetization
  $m$. In $D=1$ explicit, approximation-free evaluations of the double
  occupation $d(g,n,m)$ and the momentum distribution
  $n_{k\sigma}(g,n,m)$ were made possible by exploiting (i) relations
  for the Feynman diagrams for $d$ and $n_{k\sigma}$, derived {}from
  canonical spin and particle-hole transformations, (ii) the
  polynomial form of the diagrams in powers of $k$, $n$, and $m$, and
  (iii) an analysis of the contributions of normal and Umklapp
  processes.  In this way the calculation of $d(g,n,m)$ and
  $n_{k\sigma}(g,n,m)$ was reduced to that for $m$ $=$ $0$.
  Furthermore, new closed expressions for the momentum distribution
  $n_{k\sigma}$ were derived, facilitating numerical evaluation.
  
  The functions $d(g,n,m)$ and $n_{k\sigma}(g,n,m)$ in $D=1$ are
  qualitatively similar to those for $m$ $=$ $0$. The discontinuity
  $q_{\sigma}(g,n,m)$ of the momentum distribution at the Fermi energy
  was also calculated explicitly. It is always finite, except for the
  half-filled band without double occupation ($g$ $=$ $0$, $n$ $=$
  $1$, $m$ $=$ $0$) in which case the electrons are trivially
  localized. In all other cases the GWF describes a ferromagnetic
  Fermi liquid.
  
  Analysis of the Gutzwiller variational energy for the Hubbard chain
  with nearest-neighbor hopping shows that the GWF predicts a fully
  polarized ferromagnetic state at large enough $U$ and away from
  half-filling, in contrast to the Lieb-Mattis theorem.\cite{Lieb62a}
  This exemplifies once more the peculiarities of the GWF which
  controls correlations between the electrons only globally through
  the local Hubbard interaction.  While the GWF is an excellent wave
  function for the one-dimensional \emph{Heisenberg} model (at least
  for $m$ $=$ $0$),\cite{Gebhard87a+88a} since this only involves spin
  correlations between localized spins, it is not a very good wave
  function for the one-dimensional \emph{Hubbard} model at large $U$
  and $n\neq1$ since it does not describe density correlations well in
  this case.\cite{Vollhardt90a} As a consequence the GWF cannot
  reproduce all characteristics of the one-dimensional system.  This
  is also apparent {}from the finite discontinuity of the momentum
  distribution at the Fermi surface, which is, in fact, continuous for
  Luttinger liquids such as the one-dimensional Hubbard model.  On the
  other hand the ferromagnetic GWF represents a trial state for
  partially polarized, itinerant electrons and may thus be regarded as
  an effective, non-perturbative description of a ferromagnetic Fermi
  liquid.
  
  In view of the considerable technical complications involved in the
  present calculations it is not clear whether it will be possible to
  compute correlation functions with the GWF for $m\neq0$. Since the
  calculation of the spin-spin correlation
  function\cite{Gebhard87a+88a} for $m$ $=$ $0$ helped to gain
  considerable insight into the properties of Heisenberg-type
  models,\cite{Haldane88a,Shastry88a} a corresponding result for
  $m\neq0$ would be helpful for a better understanding of
  one-dimensional Heisenberg models in a magnetic field.

  \begin{acknowledgments}
    M.~K.\ would like to acknowledge very helpful discussions with
    G.~S.~Uhrig in the early stages of this research.  This work was
    supported in part by the Sonderforschungsbereich~484 of the
    Deutsche Forschungsgemeinschaft.  Part of this work was carried
    out while M.~K.\ was at Yale University, supported by DFG
    Grant~KO~2056 and US~NSF Grant~DMR~00-98226.
  \end{acknowledgments}

  \appendix
  \section{Calculation of polynomials for the non-mag\-net\-ic case}
  \label{app:polynomials}
  
  In this appendix we describe the derivation of closed expressions
  for the polynomials $R_p(k)$ and $Q_p(k)$ that appear in
  Sec.~\ref{sec:evaluation}.  By eliminating $Q_p(k)$ {}from MV's
  recursion formulas we obtain
  \begin{align}
    p^2\,R_p(k)+p(p+1)\,R_{p+1}(k)
    &=
    k\left((2p-1)\,R_p'(k)+2p\,R_{p+1}'(k)\right)
    \nonumber\\&~~~~~
    -(k^2-\textstyle\frac{1}{16})
    \,\Big[R_p''(k)+R_{p+1}''(k)\Big]
    ,\;\;p\geq0.\,\label{rgl1}
  \end{align}
  Furthermore, the polynomials $Q_p(k)$ can be expressed in terms of
  $R_p(k)$ as
  \begin{align}
    Q_{p+1}'(k+\textstyle\frac{1}{2})
    &=
    -(2p+1)\,R_p(k)-2(p+1)\,R_{p+1}(k)
    \nonumber\\&~~~~~
    +
    (2k-\textstyle\frac{1}{2})\,
    \left(R_p'(k)+R_{p+1}'(k)\right)
    \!,\;\;p\geq0,\label{qgl1}
  \end{align}
  together with $Q_p(\frac{3}{4})$ $=$ 0. We define $R_0(k)\defeq1$,
  $Q_0(k)\defeq0$.
  
  A closed form for $R_{p}(k)$ is obtained as follows. Using
  Eqs.~(\ref{fdowninergebnis})-(\ref{fdowninergebnis2}) we can reduce
  Eq.~(\ref{rgl1}) to
  \begin{align}
    p(p+1)\,R_p(k)
    +2k\,R_p'(k)
    +(k^2-\textstyle\frac{1}{16})\,R_p''(k)
    &=
    p^2\,R_{p-1}(k)
    ,\;\;p\geq0.\,\label{rgl2}
  \end{align}
  This is essentially the differential equation of the Legendre
  polynomials except for the inhomogeneity on the right-hand side.
  {}From an expansion in Legendre polynomials $P_n(x)$ we thus
  obtain
  \begin{align}
    R_p(k)
    &=
    \sum_{j=0}^{\lfloor\frac{1}{2}p\rfloor}
    \frac{(-1)^p p!^2(4j+1)}{(p-2j)!(p+2j+1)!}
    \binom{-\frac{1}{2}}{j}^2
    P_{2j}(4k)
    ,\;\;p\geq0,\,\label{rgl3}
  \end{align}
  which after some calculation yields Eq.~(\ref{r-function}). {}From
  ${\cal R}_{0}(x,z)$ we then obtain the explicit expressions in terms
  of a (terminating) hypergeometric function,
  \begin{align}
    R_{p}(k)
    &=
    \sum_{j=0}^{\lfloor\frac{1}{2}p\rfloor}
    \frac{(-1)^p(2p-2j)!(2j)!}{4^{j+p}(p-2j)!(p-j)!j!^3}
    (16k^2-1)^j
    \label{r-explicit1}
    \\
    &=
    \binom{-\frac{1}{2}}{p}
    \;{}_{3}F_{2}\!\left(\textstyle-\frac{1}{2}p,
      -\frac{1}{2}p+\frac{1}{2},\frac{1}{2};
      \frac{1}{2}-p,1;1-16k^2\right)
    \label{r-explicit2}
    \\
    &=
    \binom{-\frac{1}{2}}{p}
    \;{}_3F_2(\textstyle
    -p,-p,\frac{1}{2};
    \frac{1}{2}-p,1;
    \frac{4k-1}{4k+1})
    \;\textstyle\left(\frac{4k+1}{2}\right)^p
    \label{r-explicit3}
    \\
    &=
    \sum_{j=0}^{p}
    \binom{p}{j}
    \binom{-\frac{1}{2}}{j}
    \binom{-\frac{1}{2}}{p-j}
    \left(\frac{1-4k}{2}\right)^j
    \left(\frac{1+4k}{2}\right)^{p-j}
    .\,\label{r-explicit4}
  \end{align}
  Furthermore an integration by parts of Eq.~(\ref{qgl1}) leads to the
  expression for ${\cal Q}_{0}(x,z)$ in terms of ${\cal R}_{j}(x,z)$
  shown in Eq.~(\ref{q-function}).
  
  By using a hypergeometric identity to rewrite
  Eq.~(\ref{r-function}) as
  \begin{align}
    {\cal R}_{0}(x,z)
    &=
    \frac{1}{\sqrt{1-z}}
    \,
    {}_2F_1\left(\textstyle
      \frac{1}{2},
      \frac{1}{2};1;
      \frac{(x^2-1)z^2}{4(1-z)}
    \right)
    \!,\;\;
    \left|\textstyle\frac{(1-x^2)z^2}{4(1-z)}\right|<1
    ,\,\label{r-function-umgeformt}
  \end{align}
  integrating term-wise w.r.t.~$x$, and again using several
  hypergeometric identities we obtain the following explicit
  expression for ${\cal R}_{j}(x,z)$,
  \begin{align}
    {\cal R}_{j}(x,z)
    &=
    \frac{2^{2j+1}j!}{(2j)!}
    \sum_{p=0}^\infty
    \frac{(1-x)^{p+j}
      \,z^{2p}
      \,\widetilde{R}_p^{(j)}(x)}{[(2-z)^2-(x z)^2]^{p+\frac{1}{2}}}
    ,\,\label{r-integral-alt}
  \end{align}
  where the $\widetilde{R}_p^{(j)}(x)$ are polynomials of degree $p$,
  \begin{align}
    \widetilde{R}_p^{(j)}(x)
    &\defeq
    \binom{-\frac{1}{2}}{p}
    \binom{-\frac{1}{2}}{p+j}
    \;{}_{3}F_{2}\!\left(
      \textstyle-p,
      \frac{1}{2},
      j;
      1+p+j,
      \frac{1}{2}+j;
      \frac{1-x}{1+x}
    \right)
    (1+x)^{p}
    .\,\label{r-integral-alt-polynomial}
  \end{align}
  The functions ${\cal R}_{j}(x,z)$ may be evaluated via the
  series~(\ref{r-integral-alt}) for not too large values of $z$.
  Alternatively, the integration in Eq.~(\ref{r-integral}) can be
  performed numerically.
  
  Finally, we note a few special values.  {}From MV's polynomial
  relations we obtain
  \begin{align}
    {\cal Q}_{0}(x,z)+{\cal Q}_{0}(-x,z)
    &=
    1-(1-z)\,{\cal R}_{0}(x,z)+\frac{1}{2}\ln(1-z)
    .\,\label{q-symmetric-part}
  \end{align}
  {}Together with Eq.~(\ref{r-function})-(\ref{w-function}) we find in
  particular
  \begin{align}
    {\cal R}_{0}(\pm1,z)
    &=
    \frac{1}{\sqrt{1-z}}
    ,\;\;\;\;
    {\cal Q}_{0}(1,z)=0
    ,\,\label{r-special}    
    \\
    {\cal Q}_{0}(-1,z)
    &=
    1-\sqrt{1-z}+\frac{1}{2}\ln(1-z)
    .\,\label{q-special}
  \end{align}
  Furthermore the sum rule for particles outside of the Fermi surface
  {[Eq.~(\ref{sumrule})]} implies
  \begin{align}
    {\cal R}_{1}(-1,z)
    &=
    2\,{\cal R}_{1}(0,z)
    =
    \frac{2}{z}\,\ln(1-z)
    ,\\
    {\cal Q}_{1}(-1,z)
    &=
    -1+\frac{z-2}{2z}\,\ln(1-z)
    .
  \end{align}

  \section{Polynomial structure of minority spin diagrams outside
    of the Fermi surface}
  \label{app:down-out-sector-structure}
  
\begin{figure}
    \centerline{\includegraphics[height=10cm]{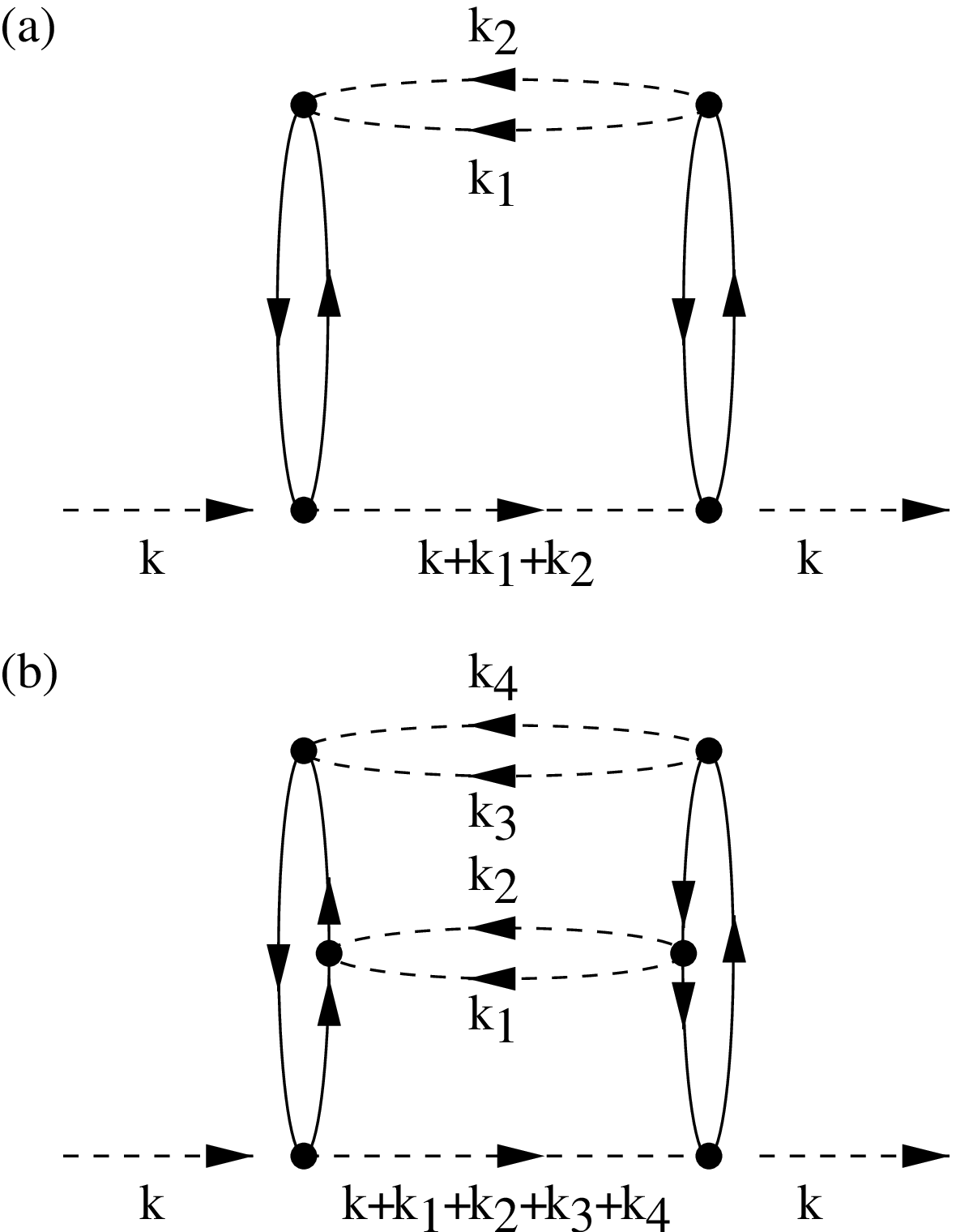}}
    \caption{A class of Feynman diagrams contributing to
      $h_{p\downarrow}^{\outside}(k,n,m)$.  Solid (broken) lines
      represent majority (minority) spins and vanish unless they carry
      momentum in the interval
      ${[-k_{\text{F}\uparrow},k_{\text{F}\uparrow}]}$
      (${[-k_{\text{F}\downarrow},k_{\text{F}\downarrow}]}$); see MV
      for diagrammatic rules. For $k>3k_{\text{F}\downarrow}$ the
      diagram in (a) vanishes since $k+k_1+k_2$ $\in$
      ${[-k_{\text{F}\downarrow},k_{\text{F}\downarrow}]}$ cannot be
      fulfilled. For $k>5k_{\text{F}\downarrow}$ the diagram in (b)
      vanishes for similar reasons, and so on for higher odd multiples
      of $k_{\text{F}\downarrow}$.\label{fig:graphs}}
  \end{figure}

  In this appendix we derive the polynomial structure of the diagrams
  $h_{p\downarrow}^{\outside}$, for which a lot of cases must be
  distinguished. As discussed at the beginning of
  Sec.~\ref{subsec:magnkocc}, different polynomials occur in the
  contribution of normal processes depending on the relation of $k$ to
  $2k_{\uparrow}-k_{\downarrow}$ and $2k_{\uparrow}+k_{\downarrow}$.
  Furthermore, due to a certain class of diagrams, shown in
  Fig.~\ref{fig:graphs}, different polynomials for
  $h_{p\downarrow}^{\outside}/n^{p}$ may in principle occur also at
  $3k_{\text{F}\downarrow}$, $5k_{\text{F}\downarrow}$, and all higher
  odd multiples of $k_{\text{F}\downarrow}$; however some
  simplification takes place, as we will show below.
  
  \emph{Case A}: $0\leq m\leq n/3$.  Here the momenta
  $2k_{\uparrow}-k_{\downarrow}$, $2k_{\uparrow}+k_{\downarrow}$,
  $3k_{\downarrow}$, etc., are ordered as follows,
  \begin{align}
    \frac{n+3m}{4}
    &\leq
    \frac{3(n-m)}{4}
    <
    \frac{3n+m}{4}
    \leq
    \frac{5(n-m)}{4}
    ,\,\label{A-ordering}
  \end{align}
  so that the contribution of normal processes to
  $h_{p\downarrow}^{\outside}$, as it appears in Eq.~(\ref{hdownout}),
  can be written in terms of three polynomials $A^{(i)}_{p}(k,m)$ of
  order $p$:
  \begin{align}
    N_p(k,0\leq m\leq{\textstyle\frac{1}{3}})
    &=
    \begin{cases}
      A^{(1)}_{p}(k,m)
      &\text{for }
      \frac{1-m}{4}<k\leq\frac{1+3m}{4}
      \\
      A^{(2)}_{p}(k,m)
      &\text{for }
      \frac{1+3m}{4}<k\leq\frac{3(1-m)}{4}
      \\
      A^{(3)}_{p}(k,m)
      &\text{for }
      \frac{3(1-m)}{4}<k\leq\frac{3+m}{4}
      \\
      0
      &\text{for }
      \frac{3+m}{4}\leq k
    \end{cases}
    \!.\,\label{N-case-A}
  \end{align}
  If $k>1-(3n+m)/4$ Umklapp processes contribute to
  $h_{p\downarrow}^{\outside}$ with momentum $1-k$, and several cases
  must be distinguished to determine the appropriate polynomials.  We
  find for $n+m\leq1$ and $m\geq3n-2$
  \begin{align}
    \frac{h_{p\downarrow}^{\outside}
      (k,n,0\leq m\leq\frac{n}{3})}{n^p}
    &=
    \begin{cases}
      A^{(1)}_{p}(\frac{k}{n},\frac{m}{n})
      &\text{for }
      \frac{n-m}{4}<k\leq\frac{n+3m}{4}
      \\
      A^{(2)}_{p}(\frac{k}{n},\frac{m}{n})
      &\text{for }
      \frac{n+3m}{4}<k\leq\frac{3(n-m)}{4}
      \\
      A^{(3)}_{p}(\frac{k}{n},\frac{m}{n})
      &\text{for }
      \frac{3(n-m)}{4}<k\leq
      \frac{1}{2}\!-\!|\frac{3n+m}{4}\!-\!\frac{1}{2}|
      \\
      A^{(3)}_{p}(\frac{k}{n},\frac{m}{n})+
      A^{(3)}_{p}(\frac{1-k}{n},\frac{m}{n})
      &\text{for }
      1\!-\!\frac{3n+m}{4}<k\leq\frac{1}{2}
      \\
      0
      &\text{for }
      \frac{3n+m}{4}<k\leq\frac{1}{2}
      \end{cases}
      \!,\,\label{hdown-a-1a}
  \end{align}
  while for $n+m\leq1$ and $m\leq3n-2$
  \begin{align}
    \frac{h_{p\downarrow}^{\outside}
      (k,n,0\leq m\leq\frac{n}{3})}{n^p}
    &=
    \begin{cases}
      A^{(1)}_{p}(\frac{k}{n},\frac{m}{n})
      &\text{for }
      \frac{n-m}{4}<k\leq\frac{n+3m}{4}
      \\
      A^{(2)}_{p}(\frac{k}{n},\frac{m}{n})
      &\text{for }
      \frac{n+3m}{4}<k\leq1\!-\!\frac{3n+m}{4}
      \\
      A^{(2)}_{p}(\frac{k}{n},\frac{m}{n})+
      A^{(3)}_{p}(\frac{1-k}{n},\frac{m}{n})
      &\text{for }
      1\!-\!\frac{3n+m}{4}<k\leq
      \frac{1}{2}\!-\!|\frac{3(n-m)}{4}\!-\!\frac{1}{2}|
      \\
      A^{(2)}_{p}(\frac{k}{n},\frac{m}{n})+
      A^{(2)}_{p}(\frac{1-k}{n},\frac{m}{n})
      &\text{for }
      1\!-\!\frac{3(n-m)}{4}<k\leq\frac{1}{2}
      \\
      A^{(3)}_{p}(\frac{k}{n},\frac{m}{n})+
      A^{(3)}_{p}(\frac{1-k}{n},\frac{m}{n})
      &\text{for }
      \frac{3(n-m)}{4}<k\leq\frac{1}{2}
    \end{cases}
    \!,\,\label{hdown-a-1b}
  \end{align}
  whereas for $n+m\geq1$
  \begin{align}
    \frac{h_{p\downarrow}^{\outside}
      (k,n,0\leq m\leq\frac{n}{3})}{n^p}
    &=
    \begin{cases}
      A^{(1)}_{p}(\frac{k}{n},\frac{m}{n})
      &\text{for }
      \frac{n-m}{4}<k\leq1\!-\!\frac{3n+m}{4}
      \\
      A^{(1)}_{p}(\frac{k}{n},\frac{m}{n})+
      A^{(3)}_{p}(\frac{1-k}{n},\frac{m}{n})
      &\text{for }
      1\!-\!\frac{3n+m}{4}<k\leq\frac{n+3m}{4}
      \\
      A^{(2)}_{p}(\frac{k}{n},\frac{m}{n})+
      A^{(3)}_{p}(\frac{1-k}{n},\frac{m}{n})
      &\text{for }
      \frac{n+3m}{4}<k\leq
      \frac{1}{2}\!-\!|\frac{3(n-m)}{4}\!-\!\frac{1}{2}|
      \\
      A^{(2)}_{p}(\frac{k}{n},\frac{m}{n})+
      A^{(2)}_{p}(\frac{1-k}{n},\frac{m}{n})
      &\text{for }
      1\!-\!\frac{3(n-m)}{4}<k\leq\frac{1}{2}
      \\
      A^{(3)}_{p}(\frac{k}{n},\frac{m}{n})+
      A^{(3)}_{p}(\frac{1-k}{n},\frac{m}{n})
      &\text{for }
      \frac{3(n-m)}{4}<k\leq\frac{1}{2}
    \end{cases}
    \!.\,\label{hdown-a-2}
  \end{align}
  
  \emph{Case B}: $n/3\leq m\leq n/2$. Here, on the other hand, we have
  the ordering
  \begin{align}
    \frac{3(n-m)}{4}
    &\leq
    \frac{n+3m}{4}
    \leq
    \frac{5(n-m)}{4}
    \leq
    \frac{3n+m}{4}
    \leq
    \frac{7(n-m)}{4},
  \end{align}
  and the contributions of normal processes to
  $h_{p\downarrow}^{\outside}$ are now given by four polynomials
  $B^{(i)}_{p}(k,m)$ of order $p$:
  \begin{align}
    N_p(k,{\textstyle\frac{1}{3}}\leq m\leq{\textstyle\frac{1}{2}})
    &=
    \begin{cases}
      B^{(1)}_{p}(k,m)
      &\text{for }
      \frac{1-m}{4}\leq k\leq\frac{1+3m}{4}
      \\
      B^{(2)}_{p}(k,m)
      &\text{for }
      \frac{3(1-m)}{4}\leq k\leq\frac{1+3m}{4}
      \\
      B^{(3)}_{p}(k,m)
      &\text{for }
      \frac{1+3m}{4}\leq k\leq\frac{5(1-m)}{4}
      \\
      B^{(4)}_{p}(k,m)
      &\text{for }
      \frac{5(1-m)}{4}\leq k\leq\frac{3+m}{4}
      \\
      0
      &\text{for }
      \frac{3+m}{4}\leq k
    \end{cases}
    \!.\,\label{N-case-B}
  \end{align}
  Let us determine the Umklapp process contributions for the region
  $1-n\leq m\leq n-1/2$,
  \begin{align}
    \frac{h_{p\downarrow}^{\outside}
      (k,n,\frac{n}{3}\leq m\leq\frac{n}{2})}{n^p}
    &=
    \begin{cases}
      B^{(1)}_{p}(\frac{k}{n},\frac{m}{n})
      &\text{for }
      \frac{n-m}{4}<k\leq1\!-\!\frac{3n+m}{4}
      \\
      B^{(1)}_{p}(\frac{k}{n},\frac{m}{n})+
      B^{(4)}_{p}(\frac{1-k}{n},\frac{m}{n})
      &\text{for }
      1\!-\!\frac{3n+m}{4}<k\leq1\!-\!\frac{5(n-m)}{4}
      \\
      B^{(1)}_{p}(\frac{k}{n},\frac{m}{n})+
      B^{(3)}_{p}(\frac{1-k}{n},\frac{m}{n})
      &\text{for }
      1\!-\!\frac{5(n-m)}{4}<k\leq\frac{3(n-m)}{4}
      \\
      B^{(2)}_{p}(\frac{k}{n},\frac{m}{n})+
      B^{(3)}_{p}(\frac{1-k}{n},\frac{m}{n})
      &\text{for }
      \frac{3(n-m)}{4}<k\leq
      \frac{1}{2}\!-\!|\frac{n+3m}{4}\!-\!\frac{1}{2}|
      \\
      B^{(3)}_{p}(\frac{k}{n},\frac{m}{n})+
      B^{(3)}_{p}(\frac{1-k}{n},\frac{m}{n})
      &\text{for }
      \frac{n+3m}{4}<k\leq\frac{1}{2}
      \\
      B^{(2)}_{p}(\frac{k}{n},\frac{m}{n})+
      B^{(2)}_{p}(\frac{1-k}{n},\frac{m}{n})
      &\text{for }
      1\!-\!\frac{n+3m}{4}<k\leq\frac{1}{2}
    \end{cases}
    \!,\,\label{hdown-b}
  \end{align}
  Now we connect the polynomials with one another via
  Eq.~(\ref{hdownrelout}), which performs the transformation $n$ $\to$
  $1$ $-$ $m$ and $m$ $\to$ $1$ $-$ $n$. This provides a link between
  Eqs.~(\ref{hdown-a-1b}) and (\ref{hdown-a-2}), the right-hand sides
  of which are thus related, line by line, via~(\ref{hdownrelout}).
  Similarly, Eqs.~(\ref{hdown-b}) and (\ref{hdown-a-1a}) are
  connected, which reveals that the distinction at momentum
  $1-5(n-m)/4$ in (\ref{hdown-b}) is in fact absent in
  Eqs.~(\ref{hdown-a-1a}). Hence we find $B^{(3)}_{p}(k,m)$ $=$
  $B^{(4)}_{p}(k,m)$, and also, by comparison with the first
  transformation, $B^{(1)}_{p}(k,m)$ $=$ $A^{(1)}_{p}(k,m)$,
  $B^{(3)}_{p}(k,m)$ $=$ $A^{(3)}_{p}(k,m)$. By inspecting the region
  $2n-1$ $\leq$ $m$ $\leq$ $1-n$ of case $B$, we find a similar
  connection to larger $m$ (i.~e., \emph{Case C}: $n/2$ $\leq$ m $\leq$
  $3n/5$), and use of Eq.~(\ref{hdownrelout}) shows that the new
  distinction at momentum $7(n-m)/4$ disappears in a similar fashion,
  and that all its polynomials likewise reduce to the ones above. It
  is not difficult to see that this simplification repeats for larger
  magnetization, i.~e., also for all $3n/5\leq m\leq n$.  We conclude
  that new polynomials at higher odd multiples of $(n-m)/4$ are ruled
  out by the symmetry of the diagrams; only the distinction at
  momentum $3(n-m)/4$ survives.  Therefore the contribution
  (\ref{hdownout}) of normal processes to $h_{p\downarrow}^{\outside}$
  can finally be written as
  \begin{align}
    N_{p}({k},{m})
    &=
    \begin{cases}
      A_{p}^{(1)}({k},{m})
      &\text{for }
      \frac{1-{m}}{4}
      < {k} \leq
      \text{min}({\textstyle{\frac{1+3{m}}{4}}},
      {\textstyle{\frac{3(1-{m})}{4}}})
      \\
      A_{p}^{(2)}({k},{m})
      &\text{for }
      m\leq{\textstyle\frac{1}{3}}
      \text{ and }
      {\textstyle\frac{1+3{m}}{4}}
      \leq {k} \leq
      {\textstyle\frac{3(1-{m})}{4}}
      \\
      B_{p}^{(2)}({k},{m})
      &\text{for }
      m\geq{\textstyle\frac{1}{3}}
      \text{ and }
      {\textstyle\frac{3(1-{m})}{4}}
      \leq {k} \leq
      {\textstyle\frac{1+3{m}}{4}}
      \\
      A_{p}^{(3)}({k},{m})
      &\text{for }
      \text{max}({\textstyle{\frac{1+3{m}}{4}}},
      {\textstyle\frac{3(1-{m})}{4}})
      \leq {k} \leq
      \frac{3+{m}}{4}
      \\
      0
      &\text{for }
      \frac{3+{m}}{4}
      \leq {k}
    \end{cases}
    \!,\,\label{downout-polynomials}
  \end{align}
  i.~e., a total of four polynomials are needed to describe
  $n_{k\downarrow}^{\outside}$; they are determined in
  Appendix~\ref{app:down-out-sector-polynomials}.

  \section{Calculation of polynomials for minority spins outside of the 
    Fermi surface}
  \label{app:down-out-sector-polynomials}
  
  In this appendix we determine the polynomials that appear on the
  right-hand side of Eq.~(\ref{downout-polynomials}).  First let us
  examine $A_{p}^{(2)}({k},{m})$. For $m$ $=$ $0$ we immediately
  obtain $A_{p}^{(2)}({k},0)$ $=$ $Q_p({k})$ by comparison with
  Eq.~(\ref{fnonmagn}).  Furthermore we can derive its behavior for
  small $m$ {}from the equation
  \begin{align}
    h_{p\downarrow}^{\outside}(k,n,m)
    &=
    h_{p\uparrow}^{\outside}(k,n,-m)
    \,,\label{h-down-h-up}
  \end{align}
  which is a simple consequence of Eq.~(\ref{negmagnkocc}). Similar to
  MV it can be shown that $h_{p\uparrow}(k,n,m)$, when regarded as a
  function of $n$, has two continuous derivatives at $n$ $=$ $1$ (for
  all $k$ $\neq$ $k_{\text{F}\uparrow}$).  Then
  Eq.~(\ref{hupinrelation}) implies that $h_{p\uparrow}(k,n,m)$ has
  the same property as a function of $m$ at $m$ $=$ $0$.  Hence the
  expression~(\ref{hupingleichung}) may be used on the right-hand side
  of Eq.~(\ref{h-down-h-up}) for small positive $m$, up to an error of
  $O(m^3)$.  Inserting the appropriate polynomials for momenta in the
  interval $(n+3m)/4<k<\text{min}(3(n-m)/4,1-(3n+m)/4)$ we obtain
  \begin{align}
    A_{p+2}^{(2)}({k},{m})
    &=
    \sum_{r=0}^{p}
    \binom{p+1}{r+1}
    (1+{m})^{r+2}{m}^{p-r}
    Q_{r+2}({\textstyle\frac{{k}+{m}/2}{1+{m}}})
    +O({m}^3)
    ,\,\label{A2-poly}
  \end{align}
  which holds for all $k$ and small $m$.
  
  The following definitions will help keep the notation compact.  Let
  $P$ be any of $A^{(1)}$, $A^{(2)}$, $B^{(2)}$, $A^{(3)}$. In
  addition to the polynomials $P_{p}(x,y)$, we define
  $\widetilde{P}_{p}(x,y)$ $\defeq$ $P_{p}(x,y)+c_{p-1}(1,y)$, and
  introduce their generating functions
  \begin{align}
    P(x,y,z)
    &\defeq
    \sum_{p=2}^{\infty}
    P_{p}(x,y)\,(-z)^p
    ,\,\label{P-function}
    \\
    \widetilde{P}^{(\alpha)}(x,y,z)
    &\defeq
    P(x,y,z)
    -
    \frac{z\,(1-y)}{2}
    +
    \frac{1}{2}\ln\frac{1-y z}{1-z}
    \,.\,\label{P-function-tilde}
  \end{align}
  
  Now we are ready to collect the relations between
  Eqs.~(\ref{hdown-a-1b}) and (\ref{hdown-a-2}) and between
  (\ref{hdown-b}) and (\ref{hdown-a-1a}) that Eq.~(\ref{hdownrelout})
  provides {[see Appendix~\ref{app:down-out-sector-structure}]}.
  Setting $x=k/n$, $y=m/n$, and $s=1/n$ in these relation, we obtain
  (for arbitrary $s$)
  \begin{subequations}\label{A-poly-relations}
    \begin{align}
      \widetilde{A}_{p+2}^{(1)}(x,y)
      &=
      \sum_{r=0}^{p}
      \binom{p}{r}
      (s-y)^{r+2}(-s)^{p-r}
      \widetilde{A}_{r+2}^{(1)}({\textstyle\frac{x}{s-y}},
      {\textstyle\frac{s-1}{s-y}})
      ,\,\label{A1-poly-relation}
      \\
      \widetilde{A}_{p+2}^{(1)}(x,y)
      +           A_{p+2}^{(3)}(s-x,y)
      &=
      \sum_{r=0}^{p}
      \binom{p}{r}
      (s-y)^{r+2}(-s)^{p-r}
      \widetilde{A}_{r+2}^{(2)}({\textstyle\frac{x}{s-y}},
      {\textstyle\frac{s-1}{s-y}})
      ,\,\label{A2-poly-relation}
      \\
      \widetilde{B}_{p+2}^{(2)}(x,y)
      +           A_{p+2}^{(3)}(s-x,y)
      &=
      \sum_{r=0}^{p}
      \binom{p}{r}
      (s-y)^{r+2}(-s)^{p-r}
      \widetilde{A}_{r+2}^{(3)}({\textstyle\frac{x}{s-y}},
      {\textstyle\frac{s-1}{s-y}})
      ,\,\label{A3-poly-relation}
      \\
      \widetilde{B}_{p+2}^{(2)}(x,y)
      +          B_{p+2}^{(2)}(s-x,y)
      &=
      -c_{p+1}(1,y)-s\,c_{p}(1,y)
      ,\,\label{B2-poly-relation}
    \end{align}
  \end{subequations}
  and other relations that are in fact implied by these.  By repeated
  combination of Eqs.~(\ref{A-poly-relations}) we find the important
  equality
  \begin{align}
    A_{p}^{(1)}(x,y) + A_{p}^{(3)}(x,y)
    &=
    A_{p}^{(2)}(x,y) + B_{p}^{(2)}(x,y)
    .\,\label{A-poly-sum}
  \end{align}
  Furthermore, by setting $s=2x$ in Eq.~(\ref{A3-poly-relation}) we
  immediately obtain the explicit expression
  \begin{align}
    B_{p+2}^{(2)}(x,y)
    &=
    -c_{p+1}(1,y)-x\,c_{p}(1,y)
    .\,\label{A2-poly-result}
  \end{align}
  In terms of generating functions our results so far can be expressed
  as
  \begin{align}
    A^{(2)}(x,y,z)
    &=
    {\cal Q}_{0}\!\left(
      {\textstyle\frac{4x-2}{1+y}},
      {\textstyle\frac{(1+y)z}{1+y z}}
    \right)
    +
    O(y^3)
    ,\,\label{A2-function-expansion}
    \\
    \widetilde{A}^{(1)}(x,y,z)
    &=
    (1-s z)\,
    \widetilde{A}^{(1)}\!\left(
      {\textstyle\frac{x}{s-y}},
      {\textstyle\frac{s-1}{s-y}},
      {\textstyle\frac{(s-y)z}{s z-1}}
    \right)
    \!,\,\label{A1-function-relation}
    \\
    \widetilde{A}^{(1)}(x,y,z)
    +
    \widetilde{A}^{(3)}(s-x,y,z)
    &=
    (1-s z)\,
    \widetilde{A}^{(2)}\!\left(
      {\textstyle\frac{x}{s-y}},
      {\textstyle\frac{s-1}{s-y}},
      {\textstyle\frac{(s-y)z}{s z-1}}
    \right)
    \!,\,\label{A2-function-relation}
    \\
    \widetilde{B}^{(2)}(x,y,z)
    &=
    (1-s z)\,
    \widetilde{B}^{(2)}\!\left(
      {\textstyle\frac{x}{s-y}},
      {\textstyle\frac{s-1}{s-y}},
      {\textstyle\frac{(s-y)z}{s z-1}}
    \right)
    =
    \frac{x z}{2}
    \ln\frac{1-y z}{1-z}
    ,\,\label{B2-function-result}
    \\
    \widetilde{A}^{(3)}(x,y,z)
    &=
    \widetilde{B}^{(2)}(x,y,z)
    +
    (1-s z)\,
    \widetilde{A}^{(3)}\!\left({\textstyle\frac{s-x}{s-y}},
      {\textstyle\frac{s-1}{s-y}},
      {\textstyle\frac{(s-y)z}{s z-1}}\right)
    \!,\,\label{A3-function-relation}
  \end{align}  
  We proceed to determine $A^{(3)}$. First we obtain $A^{(3)}(x,0,z)$
  {}from Eq.~(\ref{A2-function-relation}) at $y=0$ by inserting the
  expansion~(\ref{A2-function-expansion}) and differentiating with
  respect to $s$; we then set $s=1$, so that the error term vanishes.
  This yields an expression for $dA^{(3)}(x,0,z)/dx$, which we
  integrate with respect to $x$, using $A^{(3)}(3/4,0,z)$ $=$ $0$.
  This result for $A^{(3)}(x,0,z)$ is used in
  Eq.~(\ref{A3-function-relation}) with $s=1$, which yields
  $A^{(3)}(x,y,z)$ for arbitrary $y$,
  \begin{align}
    A^{(3)}(x,y,z)
    &=
    \frac{(1-y z)\,(4x-1-3y)}{2\,(1-y)}
    {\cal Q}_{0}\!\left(
      \textstyle
      \frac{4x-2-2y}{1-y},
      \frac{(1-y)z}{1-y z}
    \right)
    \nonumber\\&~~~~~
    -
    \frac{(1-3y)z+2}{4}    
    {\cal Q}_{1}\!\left(
      \textstyle
      \frac{4x-2-2y}{1-y},
      \frac{(1-y)z}{1-y z}
    \right)
    \nonumber\\&~~~~~
    -
    \frac{(1-y)(1-z)z}{2(1-y z)}    
    \dot{\cal Q}_{1}\!\left(
      \textstyle
      \frac{4x-2-2y}{1-y},
      \frac{(1-y)z}{1-y z}
    \right)
    \!.\,\label{A3-function-result}
  \end{align}
  The dot indicates derivative with respect to second argument.  Next
  we derive $A^{(1)}$. First we obtain $A^{(1)}(x,0,z)$ {}from
  Eq.~(\ref{A2-function-relation}) at $s=1$ and $y=0$, using
  (\ref{A2-function-expansion}) at $m=0$. This result for
  $A^{(1)}(x,0,z)$ is used in Eq.~(\ref{A1-function-relation}) with
  $s=1$, which thus yields $A^{(1)}(x,y,z)$ for arbitrary $y$,
  \begin{align}
    A^{(1)}(x,y,z)
    &=
    \frac{(1-y z)\,(3-3y-4x)}{2\,(1-y)}
    {\cal Q}_{0}\!\left(
      \textstyle
      \frac{4x-2+2y}{1-y},
      \frac{(1-y)z}{1-y z}
    \right)
    \nonumber\\&~~~~~
    +
    \frac{(1-3y)z+2}{4}    
    {\cal Q}_{1}\!\left(
      \textstyle
      \frac{4x-2+2y}{1-y},
      \frac{(1-y)z}{1-y z}
    \right)
    \nonumber\\&~~~~~
    +
    \frac{(1-y)(1-z)z}{2(1-y z)}    
    \dot{\cal Q}_{1}\!\left(
      \textstyle
      \frac{4x-2+2y}{1-y},
      \frac{(1-y)z}{1-y z}
    \right)
    \nonumber\\&~~~~~
    +
    B^{(2)}(x,y,z)
    .\,\label{A1-function-result}
  \end{align}
  Finally, $A^{(2)}(x,y,z)$ is obtained {}from Eq.~(\ref{A-poly-sum}).
  These results can be rearranged into
  Eqs.~(\ref{downoutresult})-(\ref{N-functions}).

\newpage

\begin{thebibliography}{99}
\bibitem{note-on-q} 
    Note that for non-isotropic Fermi surfaces, in particular in
    dimension $D>1$, $q_{\sigma}$ may depend not only on the magnitude
    but also on the direction of $\bm{k}_{\text{F}\sigma}$.
\end{thebibliography}
\end{document}